\documentclass[aps,pra,twocolumn,superscriptaddress,showpacs]{revtex4}
\usepackage{epsfig}
\usepackage{amsbsy,latexsym}
\usepackage{amsmath}
\usepackage{amssymb, mathrsfs}
\usepackage[mathscr]{eucal}
\usepackage{hyperref}
\usepackage{color}
\usepackage{subfigure}
\usepackage[normalem]{ulem}
\usepackage{cancel}

\def\<{\langle}
\def\>{\rangle}

\def\qed{$\,\blacksquare$\par}

\def\ket#1{|#1\>}
\def\bra#1{\<#1|}

\newcommand{\cE}{{\cal E}}

\def\Proof{\medskip\par\noindent{\bf Proof. }}

\newcommand{\la}{\langle}
\newcommand{\ra}{\rangle}

\newcommand{\mc}[1]{\mathcal{#1}}
\newcommand{\mf}[1]{\mathfrak{#1}}
\newcommand{\mb}[1]{\mathbb{#1}}
\newcommand{\mr}[1]{\mathrm{#1}}
\newcommand{\ms}[1]{\mathsf{#1}}

\newcommand{\tr}[1]{\mathrm{tr}[#1]}
\newcommand{\sis}[2]{\la #1|#2\ra}

\newcommand{\ktb}[2]{| #1\ra\la #2|}
\newcommand{\hil}{\mc H}
\newcommand{\f}{\varphi}

\newcommand{\eps}{\varepsilon}
\newcommand{\Om}{\Omega}

\newtheorem{lemma}{Lemma}
\newtheorem{proposition}{Proposition}

\newtheorem{remark}{Remark}

\newcommand{\leqc}{\leq_C}
\newcommand{\nleqc}{\nleq_C}

\begin{document}
\title{Distance to boundary and minimum-error discrimination}
\author{Erkka Haapasalo}
\affiliation{Turku Centre for Quantum Physics,~Department of Physics and Astronomy,~University of Turku,~FI-20014 Turku,~Finland}
%\author{Teiko Heinosaari}
%\affiliation{Turku Centre for Quantum Physics,~Department of Physics and Astronomy,~University of Turku,~FI-20014 Turku,~Finland}
\author{Michal Sedl\'ak}
\affiliation{Department of Optics, Palack\'y University, 17. listopadu 1192/12, CZ-77146 Olomouc, Czech Republic }
\affiliation{RCQI,~Institute of Physics,~Slovak Academy of Sciences,~D\'ubravsk\'a cesta 9,~84511 Bratislava,~Slovakia}
\author{M\'ario Ziman}
\affiliation{RCQI,~Institute of Physics,~Slovak Academy of Sciences,~D\'ubravsk\'a cesta 9,~84511 Bratislava,~Slovakia}
\affiliation{Faculty of Informatics,~Masaryk University,~Botanick\'a 68a,~60200 Brno,~Czech Republic}
\begin{abstract}
We introduce the concept of boundariness capturing the most efficient way
of expressing a given element of a convex set as a probability
mixture of its boundary elements. In other words, this number
measures (without the need of any explicit topology) how far the given element
is from the boundary. It is shown that one of the elements from the
boundary can be always chosen to be an extremal element.
We focus on evaluation of this quantity for quantum sets of states,
channels and observables. We show that boundariness is intimately
related to (semi)norms that provide an operational
interpretation of this quantity. In particular, the minimum error
probability for discrimination of a pair of quantum devices is lower
bounded by the boundariness of each of them. We proved that
for states and observables this bound is saturated and conjectured this
feature for channels. The boundariness is zero for infinite-dimensional
quantum objects as in this case all the elements are boundary elements.
\end{abstract}
\pacs{3.67.-a}
\maketitle

\section{Introduction}
The experimetal ability to switch randomly between physical apparatuses
of the same type naturally endows mathematical representatives of
physical objects with a convex structure. This makes the convexity
(and intimately related concept of probability) one of the key
mathematical features of any physical theory. Even more,
the particular ``convexity flavor'' plays a crucial role in the differences
not only between the types of physical objects, but also between the theories.
For example, the existence of non-unique convex decomposition of density
operators is the property distinguishing quantum theory from the classical one \cite{Holevo}.

Our goal is to study the convex structures that naturally appear in the quantum theory and to illustrate the operational meaning of the concepts directly linked to the convex structure. However, most of our findings are applicable for any convex set. The main goal of this paper is to introduce and investigate the concept of boundariness quantifying how far the individual elements of the convex set are from its boundary. Intuitively, the boundariness determines the most non-uniform (binary) convex decomposition into boundary elements, hence, it quantifies how mixed the element is. We will show that this concept is operationally related to specification of the most distinguishable element (in a sense of minimum-error discrimination probability). For instance, for states the evaluation of boundariness coincides with the specification of the best distinghuishable state from the given one, hence it is proportional to trace-distance \cite{Helstrom}.

The paper is organized as follows:
Section II introduces the concept of boundariness and related results
in general convex sets, the boundariness for quantum sets is evaluated
in Section III and the relation to minimum-error discrimination is described
in Section IV. Section V shortly summarizes the main results. The appendices
contain mathematical details concerning the properties of weight function,
characterization of the boundary elements of all considered quantum sets
and numerical details of the case study.

\section{Convex structure and boundariness}
\label{sect:generalZ}
In any convex set $Z$ we may define a convex preorder $\leqc$. We say $x\leqc y$ if $x$ may appear in the convex decomposition of $y$ with a non zero weight, i.e. there exist
$z\in Z$ such that $y=t x +(1-t)z$ with $0<t\leq 1$. If $x\leqc y$, then $y$ has $x$ in its convex decomposition, hence, (losely speaking) $y$ is ``more'' mixed than $x$. The
value of $t$ (optimized over $z$) can be used to quantify this relation. Namely, for any element $y\in Z$ we define the {\it weight function} $t_y:Z \to [0,1]$ assigning for every
$x\in Z$ the supremum of possible weights $t$ of the point $x$ in the convex
decomposition of $y$, i.e.
$$
t_y(x)=\sup\Big\{0\leq t < 1\,\Big|\,z=\frac{y-tx}{1-t}\in Z\Big\}\,.
$$

\begin{figure}[t]
\includegraphics[width=0.6\linewidth]{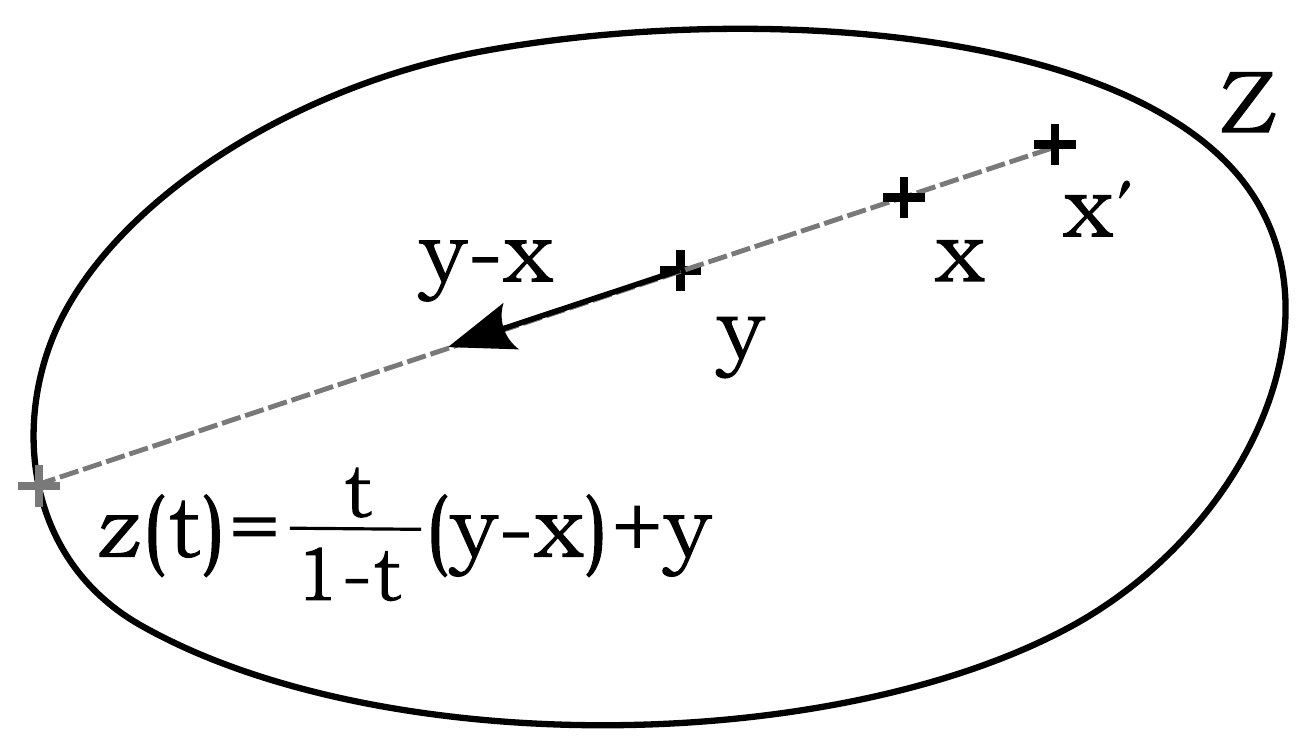}
\caption{Illustration of elements $z$ and $x^\prime$ emerging in the definition of the weight function $t_y(x)$ and in the property $t_y(x^\prime)\leq t_y(x)$, respectively.}
\label{fig:ilfdefty}
\end{figure}

Obviously, $t_y(y)=1$ and $t_y(x)=0$ whenever $x \nleqc y$. In order to understand the geometry of the optimal $z$ for a given pair of elements $x,y$, it is useful to express the
element $z$ in the form $z=y+\frac{t}{1-t}(y-x)$. As $t$ increases, $z$ moves in the direction of $y-x$ until (for value $t=t_y(x)$) it leaves the set $Z$ (see Fig. \ref{fig:ilfdefty}
for illustration). If the element $z$ associated with $t_y(x)$ is an element of $Z$, then it can be identified as a boundary element of $Z$.
The (algebraic) boundary $\partial Z$ contains all elements $y$ for which there exists $x$ such that $x\nleqc y$ (let us stress this is equivalent with the definition used in Ref.~\cite{dlp}).
Hence, for each boundary element $y$ the weight function $t_y(x)=0$ for some $x$
and also the opposite claim holds, i.e.,\ if there exists $x\in Z: t_y(x)=0$ then $y\in \partial Z$. As a consequence, $t_y(x)>0$ $\forall x\in Z$ for all inner points $y\in Z\setminus \partial Z$.

\begin{figure}[b]
 \centering
\subfigure{%
\includegraphics[width=0.3\linewidth]{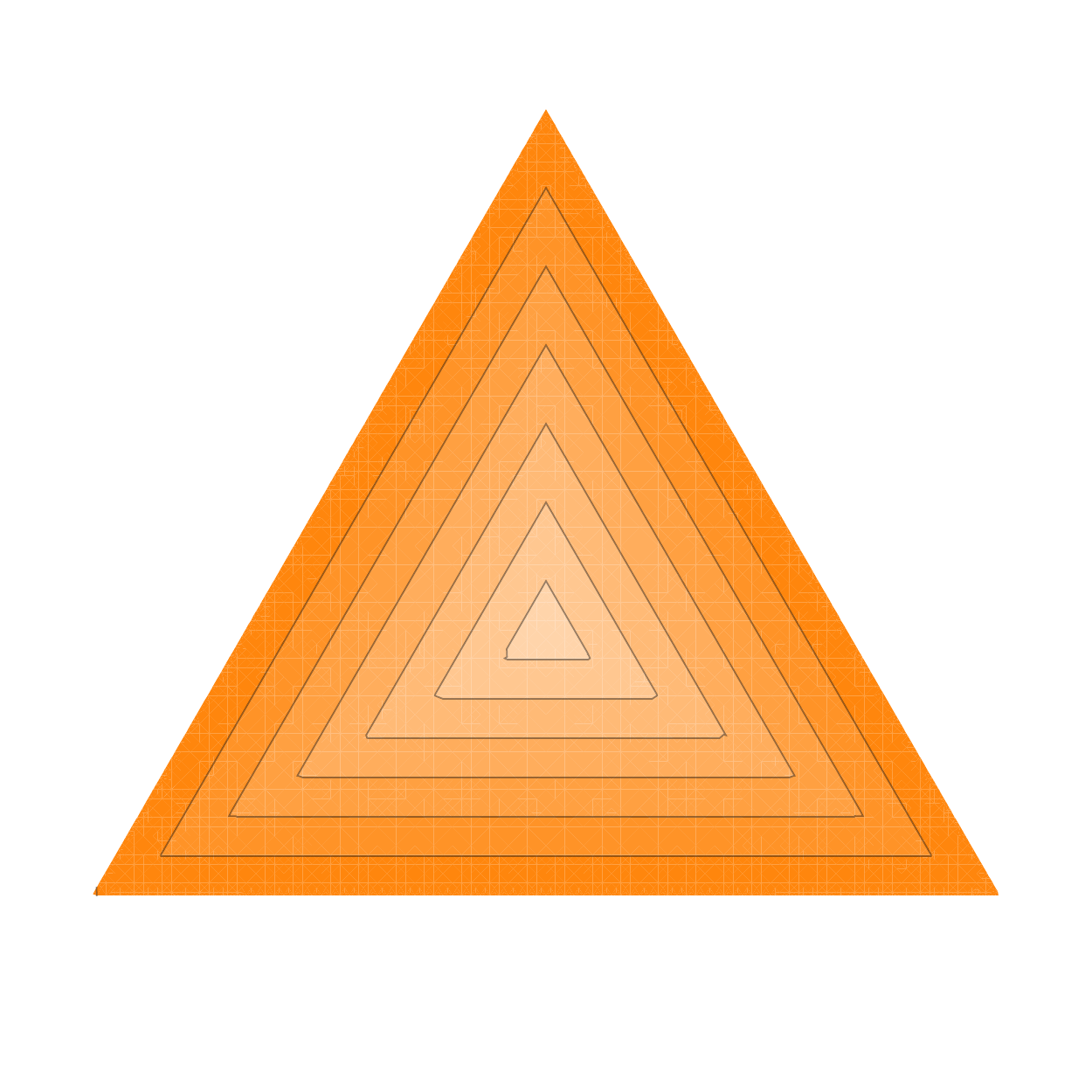}
\label{fig:subfigure1}}
\subfigure{%
\includegraphics[width=0.3\linewidth]{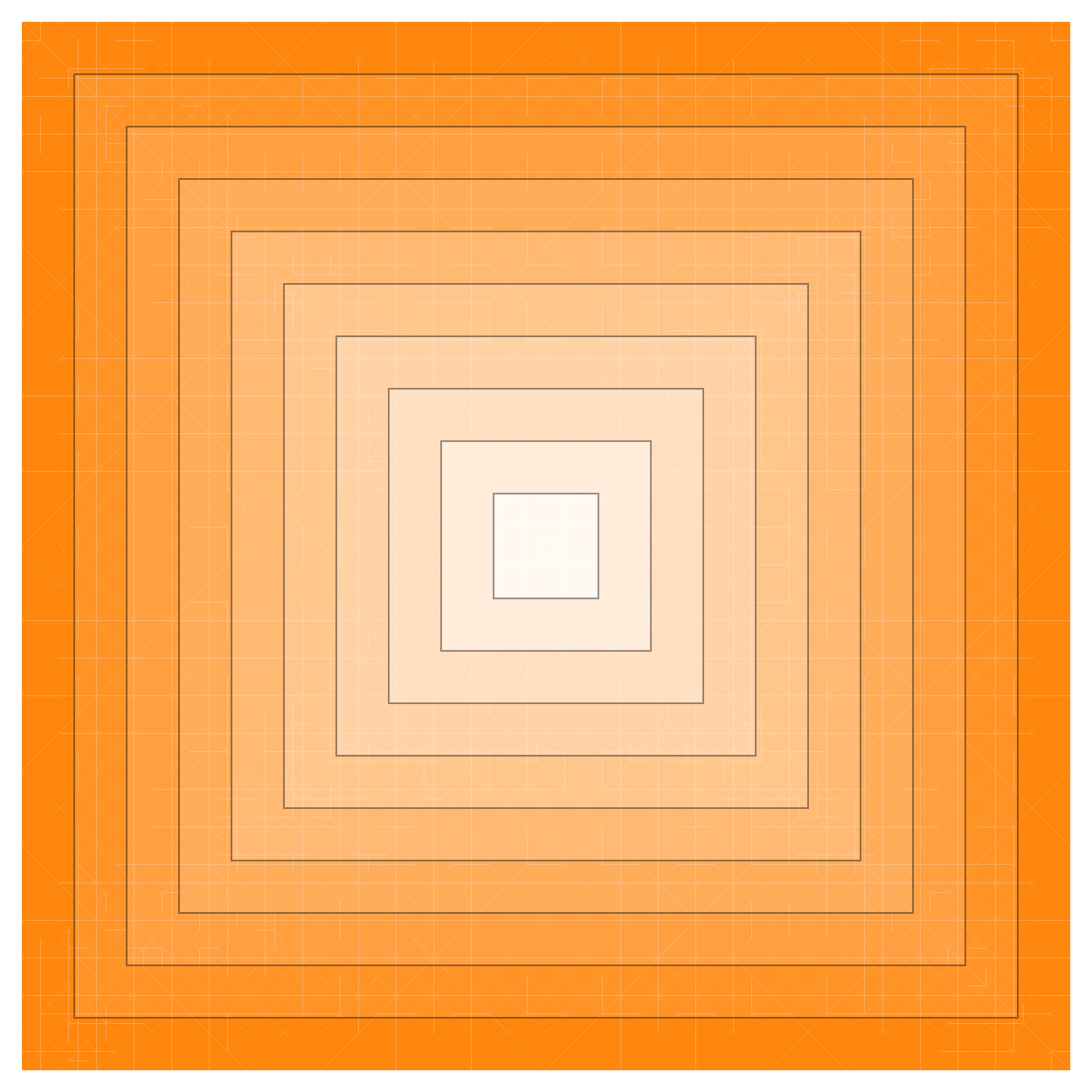}
\label{fig:subfigure2}}
\subfigure{%
\includegraphics[width=0.3\linewidth]{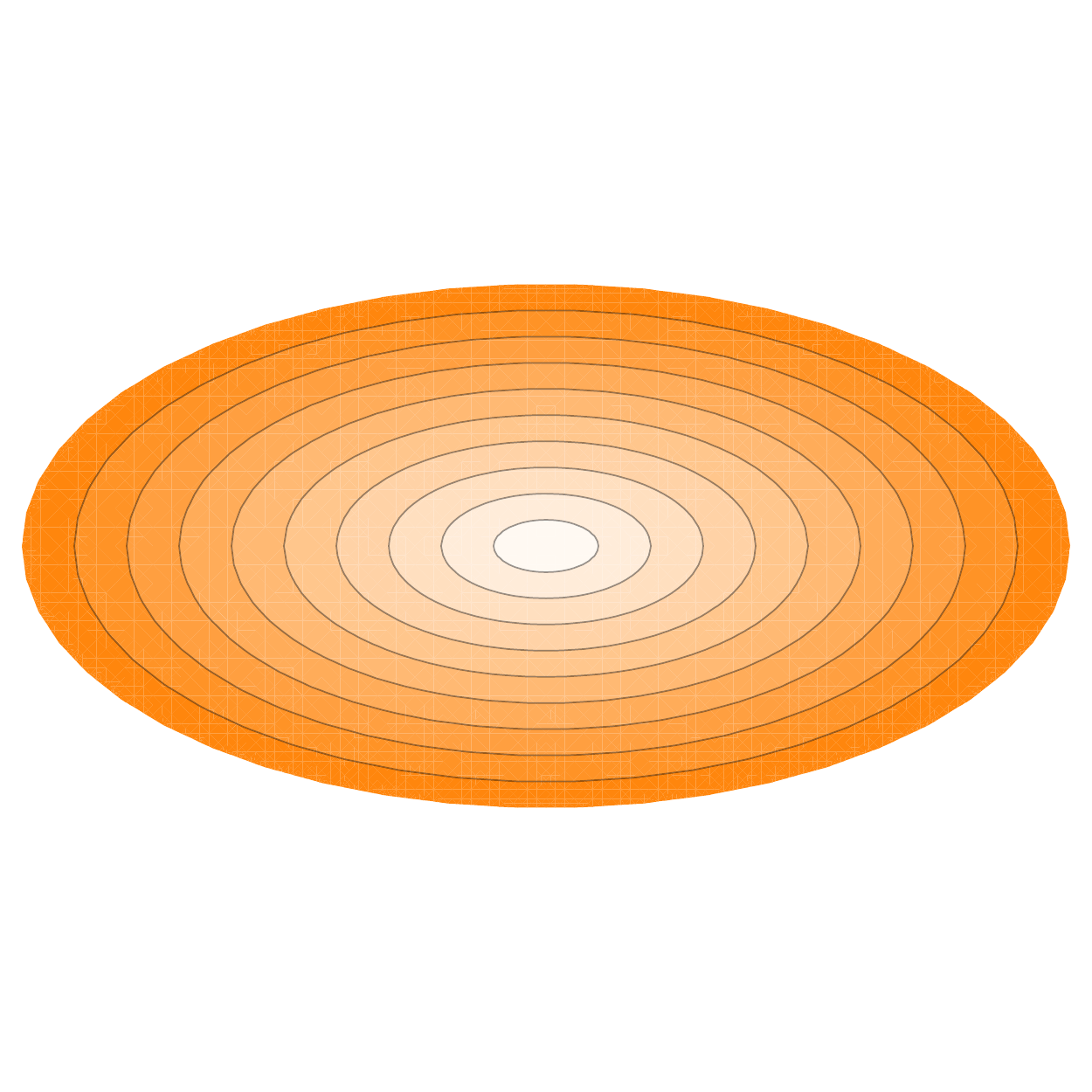}
\label{fig:subfigure3}}
\caption{(Color online) Contour plots of boundariness for simple convex sets. Let us note that the maximal value of boundariness is not the same in all of them.}
\label{fig:figure}
\end{figure}

This motivates the following definition of \emph{boundariness}
$$
b(y)=\inf_{x\in Z} t_y(x)
$$
measuring how far the given element of $Z$ is from the boundary $\partial Z$.
Suppose $x^\prime$ belongs to the line generated by $x$ and $y$, i.e. $x^\prime = y-k(y-x)$ ($x^\prime=x$ for $k=1$ and $x^\prime=y$ for $k=0$). Then
$t_y(x^\prime)\leq t_y(x)$ whenever $k\geq 1$ (see Fig. \ref{fig:ilfdefty}). Hence, the infimum can be approximated again by some boundary element of $Z$. In other words, the
value of boundariness is determined by the most non-uniform convex decomposition of $y$ into boundary elements of $Z$, i.e. $y$ can be, in a sense, approximated by expressions
$b(y)x+(1-b(y))z$ with $x,z\in\partial Z$. Therefore, $b(y)\leq 1/2$.
See Fig. \ref{fig:figure} for illustration of boundariness for simple convex sets.

\begin{lemma}\label{lemma:tconvex}
Let $y\in Z$. The inverse $x\mapsto 1/t_y(x)$ of the weight function $t_y$ is convex, i.e.,
\begin{align}
\label{eq:tconvex}
\frac{1}{t_y\big(sx_1+(1-s)x_2\big)}\leq\frac{s}{t_y(x_1)}+\frac{1-s}{t_y(x_2)} \nonumber
\end{align}
for all $x_1,\,x_2\leqc y$ and $0\leq s\leq 1$.
\end{lemma}

\begin{figure}[t]
\includegraphics[width=0.6\linewidth]{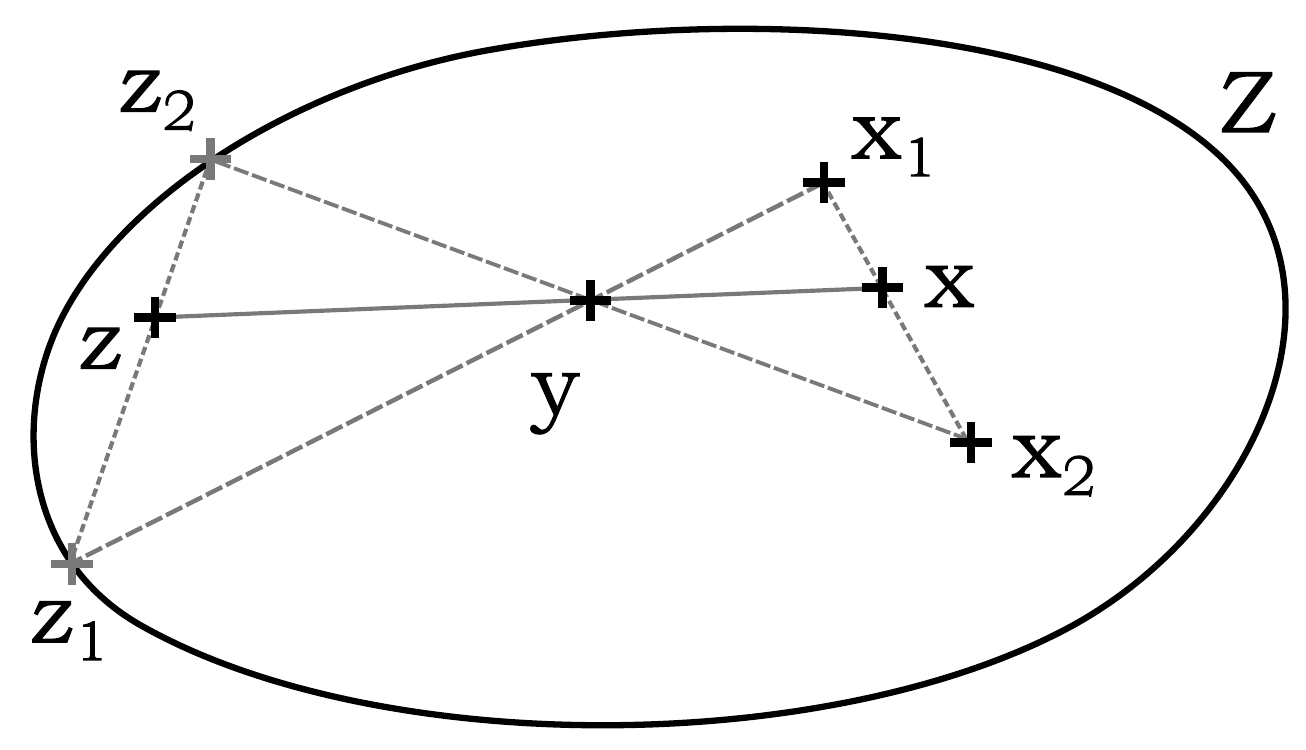}
\caption{Illustration of the proof of the Lemma \ref{lemma:tconvex}}
\label{fig:ilflemma1}
\end{figure}

\Proof
For every $0<t_i<t_y(x_i)$ $i=1,2$ we define $z_i=y-\frac{t_i}{1-t_i}(x_i-y)\in Z$. Further, we define
$x=sx_1+(1-s)x_2$ and $z=uz_1+(1-u)z_2$, where $x,\,z\in Z$ because $s\in[0,1]$ and
\begin{equation}
u=\frac{s\frac{1-t_1}{t_1}}{s\frac{1-t_1}{t_1}+(1-s)\frac{1-t_2}{t_2}}\;\in[0,1].
\end{equation}
See Fig. \ref{fig:ilflemma1} for illustration.
Straightforward calculation shows that we may write $y=tx+(1-t)z$, where $t^{-1}=st_1^{-1}+(1-s)t_2^{-1}$.
From the definition of the weight function, we have $t\leq t_y(x)$. Since this holds for all $0<t_i<t_y(x_i)$ $i=1,2$ we get
$(\frac{s}{t_y(x_1)}+\frac{1-s}{t_y(x_2)})^{-1}\leq t_y(x)$, which concludes the proof.
\qed

The following proposition is one of the key results of this section. It guarantees that one of the elements of the optimal decomposition (determining the boundariness) can be chosen
to be an extreme point of $Z$. %It is well known that any convex function defined in a finite-dimensional space is continuous on the relative interior of its domain. Although, whenever
%$y\in Z\setminus\partial Z$, one has $t_y(x)>0$ for all $x\in Z$, so that $x\mapsto1/t_y(x)$ is well defined convex function on $Z$, one
It is shown in Appendix \ref{sec:appropty} that, whenever $Z\subset\mb R^n$ for some $n\in\mb N$, the weight function $t_y$ is continuous if (and only if)
$y\in Z\setminus\partial Z$. Continuity of $t_y$ is studied in the appendices also in a slightly more general context.

\begin{proposition}\label{prop:extremal_deco}
Suppose that $Z\subset\mb R^n$ is convex and compact set. For every $y\in Z\setminus\partial Z$ there exists an extreme point $x\in Z$ such that $b(y)=t_y(x)$.
\end{proposition}

\Proof
The continuity implies that $t_y$ acquires its lowest value on
the compact set $Z$, i.e.,\ $b(y)=\inf_{x\in Z}t_y(x)=\min_{x\in Z}t_y(x)$.
Since $y\in Z\setminus\partial Z$, we have $t_y(x)>0$.
Moreover, because of the convexity of $x\mapsto 1/t_y(x)$ proven in Lemma \ref{lemma:tconvex},
it follows that
\begin{eqnarray}
\nonumber
\min_{x\in Z}t_y(x)&=&\big(\max_{x\in Z}1/t_y(x)\big)^{-1}\\\nonumber
&=&\big(\max_{x\in\mr{ext}\,Z}1/t_y(x)\big)^{-1}=\min_{x\in\mr{ext}\,Z}t_y(x)\,,
\end{eqnarray}
where $\mr{ext}\,Z$ denotes the set of extreme points of $Z$.
\qed

The convex sets appearing in quantum theory are typically compact and convex subsets of $\mb R^n$, meaning that the above proposition is applicable in our
subsequent analysis. It is easy to show that, in the context of Proposition \ref{prop:extremal_deco}, for any $y\in Z\setminus\partial Z$ and $x\in Z$ there exists an element
$z\in\partial Z$ such that $y=t_y(x)x+\big(1-t_y(x)\big)z$. This, combined with Proposition \ref{prop:extremal_deco}, yields that for any $y\in Z\setminus\partial Z$ there is
$x\in\mr{ext}\,Z$ and $z\in\partial Z$ such that $y=b(y)x+\big(1-b(y)\big)z$ when $Z$ is a convex and compact subset of $\mb R^n$.

Suppose that $y\in Z\setminus\partial Z$, where $Z\subset\mb R^n$ is a convex and compact set. Let $x\in\mr{ext}\,Z$ be an element, whose existence is guaranteed by
Proposition \ref{prop:extremal_deco}, such that $b(y)=t_y(x)$. If one had $b(y)=0$, this would mean that $t_y(x)=0$ implying that $x$ does not appear in any convex
decomposition of $y$. This yields the counterfactual result $y\in\partial Z$. Hence, $b(y)>0$ for any non-boundary element $y\in Z$, and we see that, in the context of Proposition
\ref{prop:extremal_deco}, $b(y)=0$ if and only if $y\in\partial Z$. Compactness is an essential requirement for this property. Consider, e.g.,\ a convex set $Z\subset\mb R^n$ that
has a direction, i.e.,\ there is a vector $v\in\mb R^n$ and a point $y\in Z\setminus\partial Z$ such that $y+\alpha v\in Z$ for all $\alpha>0$. Such set is not compact and one easily
sees that $b(y)=0$.

\begin{remark}{\rm (Evaluation of boundariness)}
\label{prop:algorithm}
\newline
In practise, it is useful to think about some numerical way how to evaluate
the boundariness. It follows from the definition of boundariness
that for any element $y\in Z$ written as a convex combination $y=tx+(1-t)z$ with $z\in\partial Z$
the value of $t$ (being $t_y(x)$ in this case) provides an upper bound on the
boundariness, hence $t\equiv t_y(x)\geq b(y)$. Suppose we are
given $y$ and choose some value of $t$. Recall that for a fixed $y\in Z$
and for every $x\in Z$ the element $z_t(x)=(y-tx)/(1-t)$ leaves the set $Z$
for $t=t_y(x)$. Therefore, if we choose $t\leq b(y)$ implying $t\leq t_y(x)$,
then $z_t(x)\in Z$ for all $x\in Z$. However, if it happens that $t>b(y)$,
then for some $x$ we find $t>t_y(x)$ and consequently $z_t(x)\notin Z$. Even more, according to
Proposition \ref{prop:extremal_deco} such $x$ (determining the element
$z_t(x)$ out of $Z$) can be chosen to be extremal. In conclusion,
if $t>b(y)$, then there exist $x\in\mr{ext}Z$ such that
$z_t(x)=(y-tx)/(1-t)\notin Z$.

This observation provides the basics of the numerical method we used to
test whether a given value of $t$ coincide with $b(y)$, or not.
In particular, for any $y$
we start with the maximal value of $t=1/2$ (if we do not have a better
estimate) and decrease it until we reach the value of $t$ for which
$z_t(x)\in Z$ for all $x\in\mr{ext}Z$. Equivalently, we may start with $t=0$
and increase its value until we find $t$ for which $z_{t+\varepsilon}(x)\notin Z$ for some
$x\in\mr{ext}Z$ and $\forall \varepsilon>0$.
\end{remark}

In what follows we will formulate a proposition that related relates the value
of boundariness to any (bounded) seminorm defined on the (real) vector
space $V$ containing the convex set $Z$.

\begin{proposition}\label{prop:normbound}
Consider a (semi)norm $p:V\to[0,\infty)$ such that $p(x)\leq a$
for all $x\in Z$ with some $a\geq 0$. Then
\begin{equation}\label{eq:normbound}
p(x-y)\leq2a\big(1-t_y(x)\big)\leq2a\big(1-b(y)\big)
\end{equation}
for all $x,\,y\in Z$.
\end{proposition}

\Proof
Pick $x,\,y\in Z$. The last inequality in (\ref{eq:normbound}) follows immediately from the definition of boundariness so we concentrate on the first inequality. If $t_y(x)=0$ then the
claim is trivial and follows from the triangle inequality for the seminorm. Let us assume that $t_y(x)>0$ and pick $t\in[0,t_y(x))$. According to the definition of the weight function,
we have $z(t)=(1-t)^{-1}(y-tx)\in Z$. It follows that $x-y=(1-t)\big(x-z(t)\big)$ yielding
\begin{eqnarray*}
p(x-y)&=&(1-t)p\big(x-z(t)\big)\leq(1-t)\big(p(x)+p\big(z(t)\big)\big)\\
&\leq&2a(1-t).
\end{eqnarray*}
As we let $t$ to approach $t_y(x)$ from below, we obtain the first inequality of (\ref{eq:normbound}).
\qed

In Section \ref{sec:reldiscr} we will employ this proposition to
relate the concept of boundariness to error rate of
minimum-error discrimination in case of quantum convex sets of states,
channels and observables. Shortly, the optimal values of error
probabilities are associated with the so-called
{\it base norms}~\cite{jencova2013,reeb_etal2011}, thus setting $p(x-y)=\|x-y\|_Z$
in Eq.~\eqref{eq:normbound} we obtain an operational meaning
of boundariness.  Let us stress that the base norm $\|x-y\|_Z$ can be introduced
only if certain conditions are met.

In particular, let us assume that the real vector space $V$ is equipped
with a {\it cone} $C\subset V$, i.e., $C$ is a convex set such that
$\alpha v\in C$ for any $v\in C$ and $\alpha\geq0$.
Moreover, we assume that $C$ is {\it pointed},
i.e.,\ $C\cap(-C)=\{0\}$ and {\it generating}, i.e.,\ $C-C=V$. Further,
suppose $Z\subset C$ is a {\it base} for $C$, i.e.,\ $Z$ is convex and
for any $v\in C$ there are unique $x\in Z$ and $\alpha\geq0$ with
$v=\alpha x$. Especially when $x\in Z$, there is no non-negative factor
$\alpha\neq 1$ such that $\alpha x\in Z$. Moreover, it follows that
$0\notin Z$.

Let us note that all quantum convex sets are bases for generating cones for
their ambient spaces. For example, the set of density operators $\mc S(\hil)$
on a Hilbert space $\hil$ is the base for the cone of
positive trace-class operators which, in turn, generates the real vector
space of selfadjoint trace-class operators. This is the natural ambient space
for $\mc S(\hil)$ rather than the entire space of selfadjoint bounded
operators, although the value for the boundariness of an individual state
does not change if the considered ambient space is larger than the space of selfadjoint
trace-class operators.

Whenever $Z$ is a base of a generating cone in $V$ one can define the
base norm $\|\cdot\|_Z:V\rightarrow [0,\infty)$. In particular, for each $v\in V$
$$
\|v\|_Z=\inf_{\lambda,\mu\geq 0}\{\lambda+\mu|v=\lambda x-\mu y
{\rm\ for\ some\ } x,y\in Z\}
$$
By definition $\|x\|_Z\leq 1$ for all $x\in Z$, hence,
according to Proposition \ref{prop:normbound}
\begin{align}\label{eq:base-norm-boundariness}
\|x-y\|_Z\leq 2(1-b(x))\,.
\end{align}

If $Z$ defines a base of a generating pointed cone in $V$ the weight function $t_y(x)$ has a relation to Hilbert's projective metric. Details of this relation are discussed in Appendix \ref{sec:apphpm}.
Since members of a base $Z$ can be seen as representatives of the projective space $\mb PV$, the projective metric also defines a way to compare elements of $Z$
%with applications to quantum information theory \cite{reeb_etal2011}.
which can be used to relate this metric to distinguishability measures \cite{reeb_etal2011}.

\section{Quantum convex sets}
There are three elementary types of quantum devices: sources (states),
measurements (observables) and transformations (channels).
They are represented by density operators, positive-operator valued measures,
and completely positive trace-preserving linear maps, respectively (for more details see for instance \cite{heinosaari12}).

\subsection{States}
Let us illustrate the concept of boundariness for the convex set of
{\it quantum states}, i.e. for the set of {\it density operators}
$$
\mc{S}(\mc{H}_d)=\{\varrho: \varrho\geq O,\tr{\varrho}=1\}\, ,
$$
where $\varrho\geq O$ stands for the positive-semidefinitness of the operator $\varrho$.
Suppose that the Hilbert space $\mc{H}_d$
is finite dimensional with the dimension $d$.
The boundariness $b(\varrho)$ determines a decomposition (it need not be unique)
of the state $\varrho$ into boundary elements $\xi$ and $\zeta$
$$
\varrho=b(\varrho)\xi+(1-b(\varrho))\zeta\,.
$$
A density operator belongs to the boundary if and only if it has
a nontrivial kernel (i.e. it has $0$ among its eigenvalues,
for details see appendix \ref{sec:appbfstates}). In other words there exists
vectors $\ket{\varphi}$ and $\ket{\psi}$ such that
$\xi\ket{\varphi}=0=\zeta\ket{\psi}$, respectively. Therefore,
\begin{align}
\lambda_{\min}\leq\bra{\psi}\varrho\ket{\psi}&=
b(\varrho)\bra{\psi}\xi\ket{\psi}\,, \nonumber \\
\lambda_{\min}\leq\bra{\varphi}\varrho\ket{\varphi}&=
[1-b(\varrho)]\bra{\varphi}\zeta\ket{\varphi}\,, \nonumber
\end{align}
where $\lambda_{\min}$ is the minimal eigenvalue of $\varrho$. Moreover, since
$\bra{\varphi}\zeta\ket{\varphi}\leq 1$ and $\bra{\psi}\xi\ket{\psi}\leq 1$
(because $\varrho\leq I$) it follows that boundariness is bounded in
the following way
\begin{align}
\label{eq:boundstates}
\lambda_{\min}\leq b(\varrho)\leq 1-\lambda_{\min}\,.
\end{align}
The upper bound in (\ref{eq:boundstates}) holds trivially,
because, in general, the boundariness is smaller or equal $1/2$.
On the other side, the tightness of the lower bound (\ref{eq:boundstates}) is exactly what we are interested in.

Based on our general consideration (Proposition~\ref{prop:extremal_deco})
we know we may choose $\xi$ to be the extremal element, i.e. a
one-dimensional projection. Set $\xi=\ket{\psi}\bra{\psi}$, where
$\ket{\psi}$ is the eigenvector of $\varrho$ associated with the
minimal eigenvalue $\lambda_{\rm min}$. Then
$$
\varrho=\lambda_{\min} \ket{\psi}\bra{\psi}+(1-\lambda_{\min})
\frac{\varrho-\lambda_{\min}\ket{\psi}\bra{\psi}}{1-\lambda_{\min}}
$$
is the convex decomposition of $\varrho$ into boundary elements
saturating the above lower bound, hence we have just proved the
following proposition.
\begin{proposition}
The boundariness of a state $\varrho$ of a finite-dimensional quantum system is given by
$$
b(\varrho)=\lambda_{\min}\,,
$$
where $\lambda_{\min}$ is the minimal eigenvalue of the density operator
$\varrho$.
\end{proposition}
Thus,\ the minimal eigenvalue possesses a direct operational interpretation
of the mixedness of the density operator. Indeed, the maximum $b(\varrho)=1/d$
is achieved only for the maximally mixed state $\varrho=\frac{1}{d}I$.
The infinite-dimensional case is somewhat trivial, because, according to Proposition \ref{prop:boundarystate} in the appendices, all infinite-dimensional states are on the boundary,
i.e.,\ $\partial\mc{S}(\mc{H}_\infty)=\mc{S}(\mc{H}_\infty)$. Consequently,
the boundariness of any state in this case is zero.

\subsection{Observables}
In quantum theory, the statistics of measurements is fully captured by
{\it quantum observables} which are mathematically represented by
{\it positive-operator valued measures} (POVM). Any observable $\ms C$ with
finite number of outcomes labeled as $1,\dots,n$ is represented
by positive operators (called effects) $C_1,\dots,C_n\in\mc{L}(\mc{H})$
such that $\sum_j C_j=I$. Suppose the system is prepared in a state $\varrho$.
Then, in the measurement of $\ms C$, the outcome $j$ occurs with probability
$p_j=\tr{\varrho C_j}$. The set of all observables with the fixed number $n$
of outcomes is clearly convex. We interpret $\ms C=t\ms A+(1-t)\ms B$
as an $n$-outcome measurement with effects $C_j=t A_j+(1-t)B_j$.

Let us concentrate on the finite-dimensional case $\hil=\hil_d$ and denote by $\sigma(\ms C)$ the union of all eigenvalues (spectra)
of all effects $C_j$ of a POVM $\ms C$ and denote by $\lambda_{\min}$ the smallest
number in $\sigma(\ms C)$. An observable $\ms C$ belongs to the boundary if and only if \cite{dlp}
$\lambda_{\min}=0$; this is also proved in appendix \ref{sec:appbfobs}. Using the same argumentation
as in the case of states we find that
\begin{align}
\label{eq:obsbineq}
\lambda_{\min}\leq b(\ms C)\,.
\end{align}

Suppose $\ket{\psi}$ is the eigenvector associated with
the eigenvalue $\lambda_{\min}$ of the effect $C_k$ for some
value of $k\in\{1,\dots,n\}$. Define an extremal (and projective) $n$-valued
observable $\ms A$ (in accordance with Proposition~\ref{prop:extremal_deco})
\begin{align}
\label{eq:observable_A}
A_j=\left\{
\begin{array}{lcl}
\ket{\psi}\bra{\psi} &\qquad& {\rm if\quad} j=k\\
I-\ket{\psi}\bra{\psi} &\qquad& {\rm for\ unique\quad} j\neq k\\
O &\qquad& {\rm otherwise}
\end{array}
\right.
\end{align}
The observable $\ms B$ with effects
\begin{align}
B_j=\frac{1}{1-\lambda_{\min}}(C_j-\lambda_{\min} A_j) \nonumber
\end{align}
belongs to the boundary, because
$$(1-\lambda_{\min})B_k\ket{\psi}=C_k\ket{\psi}-\lambda_{\min}
A_k\ket{\psi}=0\,,$$ hence $0\in\sigma(\ms B)$.
Using these two boundary elements of the set of $n$-valued
observables we may write $\ms C=\lambda_{\min}\ms A+(1-\lambda_{\min})\ms B$, hence
the lower bound \ref{eq:obsbineq}
can be saturated and we can formulate the following proposition:
\begin{proposition}
\label{prop:boundariness_observables}
Given an $n$-valued observable $\ms C$ of a finite-dimensional quantum system, the boundariness equals
$$
b(\ms C)=\lambda_{\min}\,,
$$
where $\lambda_{\min}$ is the minimal eigenvalue of all effects
$C_1,\dots,C_n$ forming the POVM of the observable $\ms C$.
\end{proposition}

\subsection{Channels}
\label{sec:channels}
Transformation of a quantum systems over some time interval is described
by a {\it quantum channel} mathematically represented as a trace-preserving completely
positive linear map. It is shown in the appendix \ref{sec:appbfchannels} that for infinite-dimensional
quantum systems the boundary of the set of channels coincide with the whole
set of channels, hence, the boundariness (just like for states)
vanishes. Therefore, we will focus on finite-dimensional quantum systems,
for which the channels can be isomorphically represented by so-called
Choi-Jamiolkowski operators. In particular, for a channel $\mc{E}$
on a $d$-dimensional quantum system its Choi-Jamiolkowski operator
is the unique positive operator $E=(\mc{E}\otimes \mc{I})(P_+)$, where
$P_+=\ket{\psi_+}\bra{\psi_+}$ and
$\ket{\psi_+}=\frac{1}{\sqrt{d}}\sum_{j=1}^d \ket{j}\otimes\ket{j}$.
By definition, $E$ belongs to a subset of density operators
on $\mc{H}_d\otimes\mc{H}_d$ satisfying the normalization
${\rm tr}_1 E=\frac{1}{d}I$, where ${\rm tr}_1$ denotes the partial
trace over the first system (on which the channel acts).

While the extremality of channels is a bit more complicated
than for the states, the boundary elements of the set of channels
can be characterized in exactly the same way as for states. In fact,
$\cE$ is a boundary element if and only if the associated Choi-Jamiolkowski
operator $E$ contains zero in its spectrum
(see Appendix \ref{sec:appbfchannels}). Given a channel $\cE$ we may
use the result (\ref{eq:boundstates}) derived for density operators
to lower bound the boundariness
\begin{align}
\label{eq:chbound}
\lambda_{\min}\leq b(\cE)\; ,
\end{align}
where $\lambda_{\min}$ is the minimal eigenvalue
of the Choi-Jamiolkowski operator $E$. However, since the structures
of extremal elements for channels and states are different, the
tightness of the lower bound (\ref{eq:chbound}) does not follow from
the consideration of states. Surprisingly, the following example shows
that this is indeed not the case.
\newline\newline
\noindent
 Case study: Erasure channels.
Consider a qubit ``erasure'' channel $\cE_p$ transforming an arbitrary
input state $\varrho$ into a fixed output state
$\xi_p=p\ket{0}\bra{0}+(1-p)\ket{1}\bra{1}$, $0<p<1/2$
inducing Choi-Jamiolkowski operator $E_p=\xi_p\otimes \frac{1}{2}I$.
In order to evaluate boundariness of the channel $\cE_p$, according to proposition \ref{prop:extremal_deco}, it suffices to inspect convex decompositions
\begin{align}
\label{eq:ccdecomp1}
E_p=t F + (1-t)G,
\end{align}
where $F$ corresponds to an extremal qubit channel, $G$ is a channel from the boundary. Our goal is to minimize the value of $t\equiv t_{\cE_p}(\mathcal{F})$ over extremal channels $\mathcal{F}$ in order to determine the value of boundariness.

The extremality conditions (linear independence of the set
$\{A_j^\dagger A_k\}_{jk}$) implies that extremal qubit channels
can be expressed via at most two Kraus operators $A_j$. Consequently,
the corresponding Choi-Jamiolkowski operators are either rank-one
(unitary channels), or rank-two operators. In what follows we will
discuss only the analysis of rank-one extremal channels, because it turns
out that they are minimizing the value of weight function
$t_{\cE_p}(\mathcal{F})$. The details concerning the analysis of
rank-two extremal channels (showing they cannot give boundariness)
are given in Appendix \ref{sec:appccchannel}.

Any qubit unitary channel $\mathcal{F}(\rho)=U\rho U^\dagger$ is represented by a Choi-Jamiolkowski operator $F=\ket{U}\bra{U}$, where $\ket{U}=\frac{1}{\sqrt{2}}
(\ket{u}\otimes\ket{0}+\ket{u^\perp}\otimes\ket{1})$ is a maximally entangled state and $\ket{u}\equiv U\ket{0}$, $\ket{u^\perp}\equiv U\ket{1}$. Our goal is to evaluate $t$
for which the operator $G$ specified in Eq.~(\ref{eq:ccdecomp1}) describes the channel $\mathcal{G}$ from the boundary. This reduces to analysis of eigenvalues of $(1-t)G$ that
reads $\{p,1-p,\frac{1}{2}(1-2t-\sqrt{D}), \frac{1}{2}(1-2t+\sqrt{D})\}$, where $D=(1-2p)^2 + 4 t^2$. It is straightforward to observe that they are all strictly positive for
$t<p(1-p)$, thus, the identity $t=p(1-p)$ defines the cases when channels $\mathcal{G}$ belong to the boundary of the set of channels independently of the particular choice of
the unitary channel $\mathcal{F}$. In conclusion, all unitary channels determine the same value of $t=p(1-p)$, hence, the boundariness of erasure channels equals
$b(\cE_p)=p(1-p)$.

\begin{figure}[t]
\includegraphics[width=0.9\linewidth]{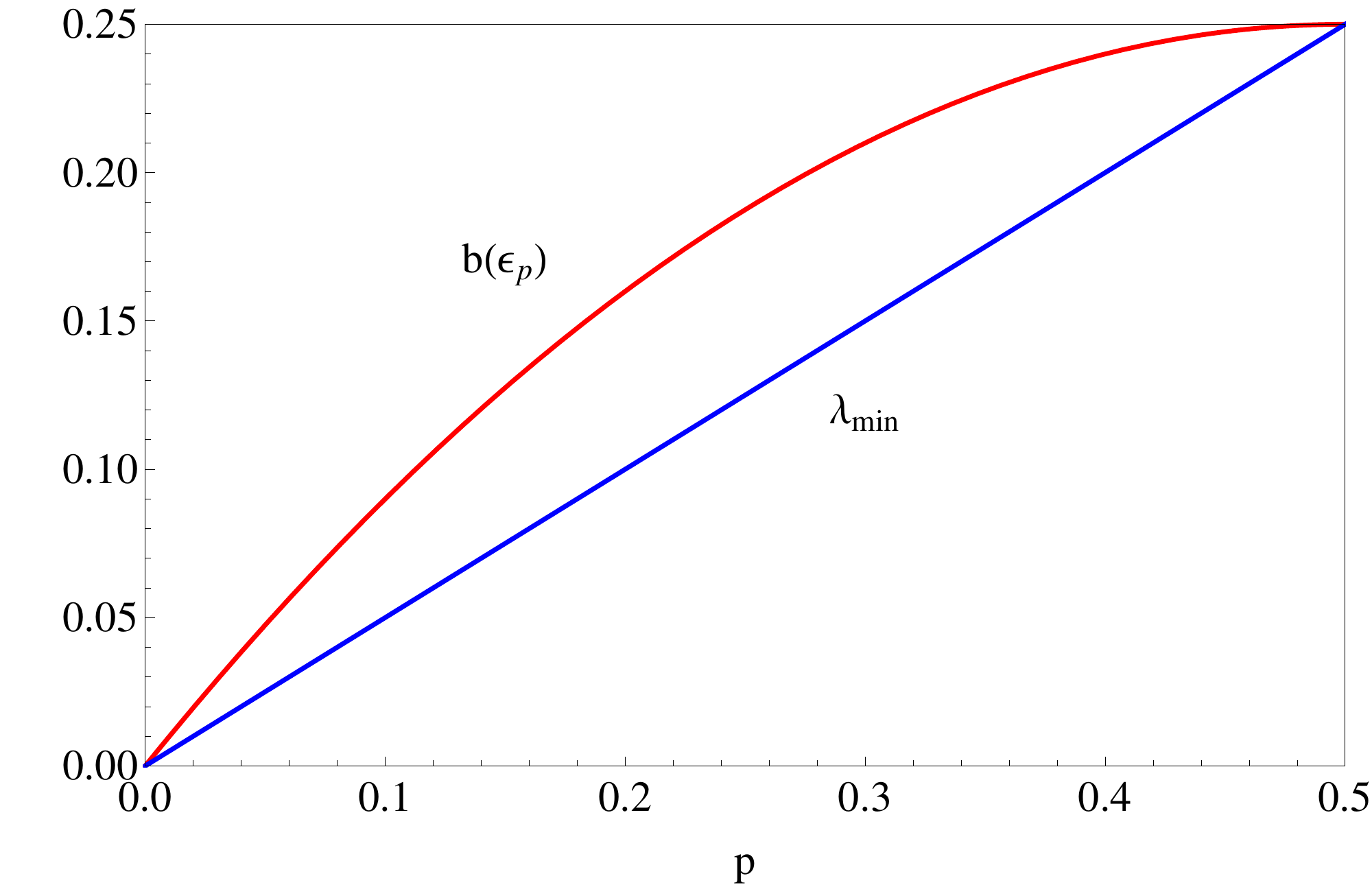}
\caption{(Color online) The strict difference between the boundariness $b$ (upper line)
and minimal eigenvalue $\lambda_{\min}$ (lower line) for erasure channels
is illustrated. Let us stress that the difference is not negligible and
it is maximal for value $p=1/4$.}
\label{fig:boundarinesscc}
\end{figure}

The example of a qubit ``erasure'' channel $\cE_p$ illustrates (see Figure \ref{fig:boundarinesscc}) that, unlike for states and observables, the boundariness of a channel $\cE$
may differ from the lower bound (\ref{eq:chbound}) given by the minimal eigenvalue of the Choi operator $E$. This finding is summarized in the following proposition.

\begin{proposition}
\label{prop:erasure}
For qubit ``erasure'' channels $\cE_p$ with $0<p<1/2$ the boundariness
is strictly larger than the minimal eigenvalue of the Choi-Jamiolkowski
operator. In particular, $b(\cE_p)=p(1-p)>\lambda_{\min}=p/2$.
\end{proposition}

Further, we will investigate for which channels (if for any)
the lower bound on boundariness is tight, i.e. when
$b(\cE_p)=\lambda_{min}$. A trivial example is provided by
channels from the boundary for which $b(\cE_p)=\lambda_{min}=0$, but
are there any other examples? Consider a channel $\cE$ such that
the minimal eigenvalue subspace of the associated Choi-Jamiolkowski
operator $E$ contains a maximally entangled state. Then a decomposition
with $t =\lambda_{min}$ exists and it corresponds to a mixture of a unitary
channel (extremal element) and some other channel from the boundary.
On the other hand, if the subspace of the minimal eigenvalue of $E$
does not contain any maximally entangled state it is natural to conjecture
that the boundariness will be strictly greater then $\lambda_{min}$.
The following proposition proves that this conjecture is valid.

\begin{proposition}
\label{prop:nolmindecomp}
Consider an inner element $\cE$ of the set of channels such that
the minimal eigenvalue subspace of its Choi-Jamiolkowski operator
$E$ does not contain any maximally entangled state. Then its boundariness
is strictly larger than the minimal eigenvalue, i.e.
$b(\mathcal{E})>\lambda_{min}$.
\end{proposition}

\Proof
We split the proof into two parts. First, we prove $t_{\cE}(\mathcal{F})>\lambda_{min}$ for any unitary channel $\mathcal{F}$ and then we prove it for any other channel
$\mathcal{F}$. Let us write the spectral decomposition of operator $E$ as
\begin{align}
E&=\sum_{k=1}^r \lambda_k P_k ,
\end{align}
where the eigenvalues $\lambda_k > 0$ are non-decreasing with $k$ (i.e. $\lambda_1=\lambda_{min}$), $P_k$ are the projectors onto eigensubspaces corresponding to $\lambda_k$ and $\sum_k P_k=I$ is the identity operator on $\mc{H}_d\otimes\mc{H}_d$. Since $\cE$ is an inner point $\lambda_1\neq 0$. The Choi-Jamiolkowski operators associated with unitary channels $\mathcal{F}$ have the form $F=\ket{\varphi}\bra{\varphi}$, where $\ket{\varphi}$ is a maximally entangled state. The assumption of the proposition implies that $P_1\ket{\varphi}\neq \ket{\varphi}$. In order to prove that $t_{\cE}(\mathcal{F})>\lambda_{min}$ it suffices to show that there exists $t>\lambda_{min}$ such that $E-tF\geq 0$ (implying $G=(E-tF)/(1-t)$ describes a quantum channel $\mathcal{G}$). It is useful to write
\begin{align}
\ket{\varphi}=\sqrt{\alpha}\ket{v}+\sqrt{1-\alpha}\ket{v_\perp},
\end{align}
where $0\leq \alpha < 1$, $P_1\ket{v}=\ket{v}$ and $P_1\ket{v_{\perp}}=0$.
Define a positive operator $X=\lambda_1 \ket{v}\bra{v} + \lambda_2 \ket{v_\perp}\bra{v_\perp}$ %with $\lambda_2>\lambda_1$
and write $E-tF=E-X+X-tF$. The operator $E-X$ is clearly positive. Further, we will show that $X-tF$ is positive
when we set $t=\lambda_1\lambda_2 / [\lambda_1+(\lambda_2-\lambda_1)\alpha] > \lambda_{\min}$ %%, hence, the operator
 and as a consequence $E-tF\geq 0$.
By definition, $X-tF$ acts nontrivially in two-dimensional subspace spanned by
vectors $\ket{v}$ and $\ket{v_\perp}$. Within this subspace it has eigenvalues
$0$ and $\lambda_2+\lambda_1-t>0$, hence, it is positive. This concludes the
first part of the proof concerning decompositions with unitary channels.

Now, let us assume that the channel $\mathcal{F}$ is not unitary.
Since the Choi-Jamiolkowski operator $F$ associated with the channel
$\mathcal{F}$ is a density operator, it follows that its maximal
eigenvalue $\mu_{max}\leq 1$ (saturated only for unitary channels).
Set $t=\lambda_{min}/ \mu_{max}$. Then for non-unitary channels
$t>\lambda_{\min}$ %%as it is required
and since
$0<\lambda_{min}\leq 1/d^2\leq\mu_{max}$ it follows that $0<t\leq 1$.
For all vectors $\ket{\varphi}$
\begin{align}
\label{eq:expGlb}
\bra{\varphi}E-t F\ket{\varphi}\geq \lambda_{min}
-\frac{\lambda_{min}}{\mu_{max}} \mu_{max}=0,
\end{align}
and, therefore, $G=(E-tF)/(1-t)\geq 0$, too. As in the first part of the proof
this means that $t_{\mathcal{E}}(\mathcal{F})>\lambda_{min}$ for all
non-unitary boundary channels $\mathcal{F}$, because we found decomposition
$\mathcal{E}=t\mathcal{F}+(1-t)\mathcal{G}$ with $t>\lambda_{min}$.
%Therefore, the strict inequality $b(\mathcal{E})>\lambda_{min}$ follows
%and the statement of the proposition holds.

The above two parts of the proof show that $t_{\mc E}(\mc F)>\lambda_{min}$ for the channel $\mc E$ of the claim and for any channel $\mc F$. The claim follows from the
observation that, according to Proposition \ref{prop:extremal_deco}, $b(\mc E)=t_{\mc E}(\mc F)$ for some (extreme) channel $\mc F$ and, especially for this optimal channel,
$t_{\mc E}(\mc F)>\lambda_{min}$.
\qed

%%%%%%%%%%%%%%%%%%%%%%%%%%%%%%%%%%%%%%%%%%%%%%%%%%%%%%%%%%%%%%%%%%%
%%%%%%%%%%%%%%%%%%%%%%%%%%%%%%%%%%%%%%%%%%%%%%%%%%%%%%%%%%%%%%%%%%%
\section{Relation to minimum-error discrimination}
\label{sec:reldiscr}
Quantum theory is known to be probabilistic, hence, individual outcomes
of experiments have typically very limited (if any) operational interpretation.
One example of this type is the question of {\it discrimination} among
a limited number of quantum devices. In its simplest form
the setting is the following. We are given an unknown quantum device, %being
which is with equal prior probability
either $A$, or $B$ ($A$ and $B$ are known to us). Our task is to design an
experiment in which we are allowed to use the given device only once
and we are asked to conclude the identity of the device.
Clearly, this cannot be done in all cases unless some imperfections
are allowed. There are various ways how to formulate the discrimination task.

The most traditional \cite{Helstrom, Holevo} one is aimed to minimize
the average probability of error of our conclusions. Surprisingly,
the success is quantified by norm-induced distances \cite{chiribella2009},
hence, the discrimination problem provides a clear operational interpretation
of these norms.
%Without specifying the norm $\|A-B\|$ explicitly, we may express the optimal error probability of minimum-error discrimination as follows
We may express the optimal error probability of minimum-error
discrimination as follows
\begin{align}
\label{eq:pevnorm}
p_{\rm error}(A,B)=\frac{1}{2}(1-\frac{1}{2}\|A-B\|)\, ,
\end{align}
where the type of the norm $\|A-B\|$ depends on the considered problem.
%{\bf where the norm $\|A-B\|$ is specified by the considered problem. }

Recently, it was shown in Ref.~\cite{jencova2013} %%\cite{reeb_etal2011}
that in general convex settings the so-called base norms are solutions to minimum-error discrimination problems. In particular, it was also shown that base norms coincide with the completely
bounded (CB) norms in the case of quantum channels, states and observables, thus, according to Proposition \ref{prop:normbound} and
Eq.~\eqref{eq:base-norm-boundariness} the following inequality
holds
$$
\|A-B\|_Z\equiv\|A-B\|_{\rm cb}\leq 2(1-b(A))\,.
$$
In rest of this section we will illustrate that for quantum structures the
base norms (being completely bounded norms) and boundariness are
intimately related. We will investigate how tight the above
inequalities are for particular quantum convex sets.

\subsection{States}
Let us start with the case of quantum states, for which
the CB norm coincides with the trace-norm (see for instance
\cite{jencova2013,chiribella2009}),
i.e. $||A||_{\rm tr}=\tr{|A|}$.
Recall that the conclusion of Proposition \ref{prop:normbound},
when applied for states, is
\begin{align}
\label{eq:prop1fstates}
\|\varrho-\xi\|_{\rm tr}\leq 2[1-b(\varrho)]\,.
\end{align}
Using the absolute scalability of the norm the roles of $\varrho$ and $\xi$ can be exchanged and from (\ref{eq:pevnorm}) and (\ref{eq:prop1fstates}) it follows that
$$
p_{\rm error}(\varrho,\xi)\geq\frac{1}{2}\max\{b(\varrho),b(\xi)\}\,,
$$
i.e. the mixedness of states measured by their boundariness lower bounds
the optimal error probability of discrimination between them. Moreover, for a given
state $\varrho$ we may write
$$
\min_\xi p_{\rm error}(\varrho,\xi) \geq \frac{1}{2}b(\varrho)\,,
$$
hence interpreting the boundariness as the limiting value of
the best distinguishability of the state $\varrho$ from any
other state. In other words, the boundariness determines
the information potential of the state as the distinguishability
of states is the key figure of merit for quantum communication
protocols \cite{wilde2013}.

As before, let $\ket{\psi}$ be the state for which
$\varrho\ket{\psi}=\lambda_{\min}\ket{\psi}$. It is straightforward to
see that
$$
\|\varrho - \ket{\psi}\bra{\psi}\|_{\rm tr}=2(1-\lambda_{\min})\,.
$$
Hence, the upper %%lower
bound (\ref{eq:prop1fstates}) can be saturated and we have proven the
following proposition.
\begin{proposition}
For a given state $\varrho$
$$
\sup_\xi \|\varrho - \xi\|_{\rm tr}=2(1-b(\varrho))\,.
$$
\end{proposition}
In particular, this implies that the states from the boundary
(with $b(\varrho)=0$) can be used as noiseless carriers of bits
of information as for each of them one can find a perfectly
distinguishable "partner" state.

\subsection{Observables}
For observables we may formulate an analoguous result:
\begin{proposition}
Suppose that $\ms C$ is an $n$-valued observable. Then
$$
\sup_{\ms A} \|\ms C-\ms A\|=2(1-b(\ms C))\,,
$$
where $\|\cdot\|$ is the base norm (identified with completely
bounded norm) for observables.
\end{proposition}

\Proof
We will prove that $\ms A$ defined in Eq. (\ref{eq:observable_A})
yields the supremum of the claim.
Let us recall that $\ket{\psi}$ (used in definition of $\ms A$)
is the vector defined by the relation $C_k\ket{\psi}=\lambda_{\min}\ket{\psi}$
for some $k$. According to Proposition \ref{prop:normbound}
\begin{align}
\label{eq:prop1fobs}
\|\ms C-\ms A\|\leq 2(1-\lambda_{\min})\,,
\end{align}
where the norm $\|\ms C-\ms A\|$ (the base norm $=$ completely bounded norm $=$ diamond norm) can be evaluated as \cite{jencova2013}
\begin{eqnarray*}
\|\ms C-\ms A\|
&=&\sup_\varrho \sum_j|\tr{\varrho(C_j-A_j)}|
\end{eqnarray*}
Assuming $\varrho=\ket{\psi}\bra{\psi}$ we obtain
$$
\|\ms C-\ms A\|\geq 1-\lambda_{\min}+\sum_{j\neq k} \bra{\psi}C_j\ket{\psi}
$$
because $\bra{\psi}A_j\ket{\psi}=0$ for $j\neq k$,
$\bra{\psi}A_k\ket{\psi}=1$ and $\bra{\psi}C_k\ket{\psi}=\lambda_{\min}$.
Moreover, since $\sum_{j\neq k}\bra{\psi}C_j\ket{\psi}=1-\bra{\psi}C_k\ket{\psi}=
1-\lambda_{\min}$ we find that for the chosen observables $\ms C$, $\ms A$ we have
$\|\ms C-\ms A\|\geq 2(1-\lambda_{\min})$. Combining this with the
lower bound (\ref{eq:prop1fobs}) valid for any observable we have proven the proposition.
\qed

\subsection{Channels}
For channels the boundariness is not given by minimal eigenvalue of
the Choi-Jamiolkowski operator. Actually, we are missing an analytical form
of channel's boundariness. %%However, even if $b(\cE)>\lambda_{\min}$
Hence, in general
the saturation of the inequality
\begin{align}
\label{eq:channelbound1}
\sup_{\mc{F}}\|\mc{E}-\mc{F}\|_{\rm cb}\leq 2(1-b(\mc{E}))
\end{align}
is open and we %%can
chose to
test the saturation of the bound for the examples
of quantum channels that we studied in Section \ref{sec:channels}. Let us
stress that analytical expressions of the completely bounded norm are
rather rare, but there exist efficient numerical methods for its
evaluation \cite{watrous2009}.

For the qubit ``erasure'' channel $\cE_p$ that transforms an arbitrary input state $\varrho$ into a fixed output state $\xi_p=p\ket{0}\bra{0}+(1-p)\ket{1}\bra{1}$ the completely
bounded norm $\|\cE_p-\mc{F}\|_{\rm cb}$ can be expressed as
\begin{align}
\|\cE_p-\mc{F}\|_{\rm cb}=\sup_{\|\psi\|=1} \|(\cE_p-\mc{F})\otimes \mathcal{I} (\ket{\psi}\bra{\psi})\|_{\rm tr} ,
\end{align}
where $\mathcal{I}$ is the qubit identity channel and $\ket{\psi}$ is a two qubit state.
Choice of $\mathcal{F}=\mathcal{I}$ and
\begin{align}
\ket{\psi}=\sqrt{1-p}\ket{0}\otimes\ket{0}+\sqrt{p}\ket{1}\otimes\ket{1}
\end{align}
lower bounds the norm in (\ref{eq:channelbound1}) by $2(1-p(1-p))$ as can be seen by direct calculation.
Due to the result $b(\cE_p)=p(1-p)$ from section \ref{sec:channels} this can be equivalently written as
$2(1-b(\cE_p))\leq \sup_{\mathcal{F}} \|\cE_p-\mc{F}\|_{\rm cb}$, which implies that the bound (\ref{eq:channelbound1}) is tight for the channel $\cE_p$.

Let us, further, consider the class of channels whose Choi operator $E$ 
contains some maximally entangled state $\ket{\phi}$ in its minimal
eigenvalue subspace. For these channels $b(\cE)=\lambda_{min}$
(see section \ref{sec:channels}). Choose $\mathcal{F}$ to be a unitary channel,
i.e. $F=\ket{\phi}\bra{\phi}$ and set
$\ket{\psi}=1/\sqrt{2}(\ket{0}\otimes\ket{0}+\ket{1}\otimes\ket{1})$ (maximally
entangled state). Then
\begin{align}
\nonumber
 \|E-F\|_{\rm tr}=\|(\cE-\mc{F})\otimes \mathcal{I} (\ket{\psi}\bra{\psi})\|_{\rm tr} \leq \|\cE-\mc{F}\|_{\rm cb},
\end{align}
and direct calculation gives $\|E-F\|_{\rm tr}=2(1-\lambda_{min})=2(1-b(\cE))$.
Altogether, we have shown
\begin{align}
 2(1-b(\cE)) \leq \sup_{\mc{F}}\|\cE-\mc{F}\|_{\rm cb},
\end{align}
which means that for this type of channels the bound
(\ref{eq:channelbound1}) is tight.

%%%%%%%%%%%%%%%%%%%%%%%%%%%%%%%%%%%%%%%%%%%%%%%%%%%%%%%%%%%%%%%%%%%%%%%%%%%%%
%%%%%%%%%%%%%%%%%%%%%%%%%%%%%%%%%%%%%%%%%%%%%%%%%%%%%%%%%%%%%%%%%%%%%%%%%%%%%
%%%%%                 SUMMARY                                          %%%%%%
%%%%%%%%%%%%%%%%%%%%%%%%%%%%%%%%%%%%%%%%%%%%%%%%%%%%%%%%%%%%%%%%%%%%%%%%%%%%%
%%%%%%%%%%%%%%%%%%%%%%%%%%%%%%%%%%%%%%%%%%%%%%%%%%%%%%%%%%%%%%%%%%%%%%%%%%%%%
\section{Summary}
Convexity is one of the main mathematical features of modern science and it is natural to ask how the physical concepts and structures are interlinked with the existing convex structure. Using only the convexity we introduced the concept of boundariness and investigated its physical meaning in statistical theories such as quantum mechanics. Intuitively, the boundariness quantifies how far an element of the convex set is from its boundary. The definition of the boundary is based solely on the convexity and no other mathematical structure of the set is assumed.

We have shown that the value of boundariness $b(y)$ identifies the most
non-uniform convex decomposition of inner element $y$ into a pair of
boundary elements. Further, we showed (Proposition \ref{prop:extremal_deco})
that for compact convex sets such optimal decomposition is achieved
when one of the boundary points is also extremal. This surprising property
simplifies significantly our analysis of quantum convex sets and allowes us
to evaluate the value of boundariness.

In particular, we have found that, %%unlike for
in contrast to the case of states
and observables, for channels the general lower bound on boundariness
($b\geq \lambda_{\min}$)  given by the minimal eigenvalue of the Choi-Jamiolkowski
representation is not saturated (see Section III). We illustrated this
feature explicitely for the class of qubit ``erasure'' channels $\cE_p$
mapping whole state space into a fixed state $\xi_p=p\ket{0}\bra{0}+(1-p)\ket{1}\bra{1}$ ($0<p<1/2$). The boundariness of this channel was found to be
$b(\cE_p)=p(1-p)>\lambda_{\min}=p/2$ (Proposition \ref{prop:erasure}). We showed that the saturation of the bound is equivalent with existence of maximally entangled state in
the minimal eigenvalue subspace of the channel's Choi-Jamiolkowski operator. Let us stress that the boundariness vanishes for infinite dimensional systems, because the associated convex
sets contain no interior points (discussed in Appendix B).

Concerning the operational meaning of boundariness, we first demonstrated that the boundariness can be used to upper bound any (semi)norm induced distance providing the
(semi)norm is bounded on the convex set. An example of such norm is the base norm which is induced solely by the convex structure of the set. Recently, it was shown in
Ref.~\cite{jencova2013} that for the sets of quantum states, measurements and evolutions base norms coincide with so-called completely bounded norms. These norms are
known \cite{chiribella2009, gutoski2012} to appear naturally in quantum minimum error discrimination tasks. As a result, this connection provides a clear operational interpretation for
the boundariness as described in Section IV.

More precisely, if we want to determine in which of the two known (equally likely) possibilities $A$ or $B$ an unknown state (or measurement, or channel) was prepared and given to us, the probability of making an erroneous conclusion exceeds one half times the boundariness for any of the elements $A$ and $B$. For a generic pair of possibilities $A$ and $B$ this bound is not necessarily tight, however if we keep $A$ fixed then the boundariness of $A$ is proportional to the minimum error probability discrimination of $A$ and the most distinguishable quantum device from $A$. To be precise this was shown only for states and observables (in which case the analytic formula for boundariness was derived), but we conjecture that this feature holds also for quantum channels. We verified this conjecture for erasure channels and the class of channels containing a  maximally entangled state in the minimum eigenvalue subspace of their Choi-Jamiolkoski operators.

In conclusion let us mention a rather intriguing observation. In all the cases
we have met the ``optimal'' decompositions (determining the value of boundariness)
contain pure states, sharp observables and unitary
channels. In other words, only special subsets of extremal elements
(for observables and channels) are needed. This is true for all states and
for all observables. The case of channels is open, but no counter-example
is known. This observation suggests that the concept of boundariness
could provide some operational meaning to sharpness of observables
and unitarity of evolution.

%%%%%%%%%%%%%%%%%%%%%%%%%%%%%%%%%%%%%%%%%%%%%%%%%%%%%%%%%%%%%%%%%%%%%%%%
%%%%%%%%%%%%%%%%%%%%%%%%%%%%%%%%%%%%%%%%%%%%%%%%%%%%%%%%%%%%%%%%%%%%%%%%
%%%%%%%                        ACKNOWLEDGMENT                     %%%%%%
%%%%%%%%%%%%%%%%%%%%%%%%%%%%%%%%%%%%%%%%%%%%%%%%%%%%%%%%%%%%%%%%%%%%%%%%
%%%%%%%%%%%%%%%%%%%%%%%%%%%%%%%%%%%%%%%%%%%%%%%%%%%%%%%%%%%%%%%%%%%%%%%%

\acknowledgments
The authors would like to thank Teiko Heinosaari for stimulating this work
and insightful discussions. This work was supported by COST Action MP1006
and  VEGA 2/0125/13 (QUICOST). E.H. acknowledges financial support from the
Alfred Kordelin foundation.
M.S. acknowledges support by the Operational Program Education for Competitiveness - European Social Fund
(project No. CZ.1.07/2.3.00/30.0004) of the Ministry of Education, Youth and Sports of the Czech Republic.
M.Z. acknowledges the
support of projects GACR P202/12/1142 and RAQUEL.

%%%%%%%%%%%%%%%%%%%%%%%%%%%%%%%%%%%%%%%%%%%%%%%%%%%%%%%%%%%%%%%%%%%%%%%%
%%%%%%%%%%%%%%%%%%%%%%%%%%%%%%%%%%%%%%%%%%%%%%%%%%%%%%%%%%%%%%%%%%%%%%%%
%%%%%%%                         APPENDICES                        %%%%%%
%%%%%%%%%%%%%%%%%%%%%%%%%%%%%%%%%%%%%%%%%%%%%%%%%%%%%%%%%%%%%%%%%%%%%%%%
%%%%%%%%%%%%%%%%%%%%%%%%%%%%%%%%%%%%%%%%%%%%%%%%%%%%%%%%%%%%%%%%%%%%%%%%
\appendix

\section{Properties of the weight function}
\label{sec:appropty}

The purpose of this appendix is to prove results that are needed for Proposition \ref{prop:extremal_deco}.
Let us first recall a few basic definitions in linear analysis. Suppose that $V$ is a real vector space.
For a subset $X\subset V$ we denote by $V_X$ the smallest affine subspace of $V$
containing $X$. For any $x\in X$, the linear subspace $V_X-x$ is just the linear hull of $X-x$,
where we introduced the notation $X-x \equiv \{y-x\,|\,y\in X\}$.
We say that $U\subset V$ is {\it absorbing} if for every $v\in V$ there
is $\alpha>0$ such that $\alpha^{-1}v\in U$; especially $0\in U$. %When $V$ is a
%topological vector space, we call the interior of $X$ within the affine space $V_X$ equipped with the relative topology induced by $V$ as the {\it relative interior of $X$}.
The following lemma gives another characterization for the boundary of a convex set $Z$, which is useful for studying the continuity properties of the weight function.

\begin{lemma}\label{prop:innerabsorb}
Suppose that $Z$ is a convex subset of a real vector space $V$. An element $y\in Z$ is inner, i.e.,\ $y\in Z\setminus\partial Z$ if and only if $Z-y$ is absorbing in the subspace
$V_Z-y$.
\end{lemma}

\Proof
Let us assume that $y$ is an inner point of $Z$ and suppose that $v\in V_Z-y$. For simplicity, let us assume that $v\neq0$. The convexity of $Z-y$ and the definition of $V_Z$ yield
that there are $d_+,\,d_- \in Z-y$ and $\lambda_+,\,\lambda_-\geq0$, where $\lambda_+>0$ or $\lambda_->0$ such that $v=\lambda_+d_+-\lambda_-d_-$. The fact that $y$ is an
inner point implies that when $d_-\in Z-y$ then $\exists q>0$ such that $-q d_-\in Z-y$. Hence, $v=\alpha d$, where $\alpha=\lambda_++\lambda_-/q>0$,
$d=\frac{\lambda_+}{\alpha}d_++\frac{\lambda_-}{q\alpha}(-q d_-)\in Z-y$, which proves that $Z-y$ is absorbing in $V_Z-y$. Suppose now that $Z-y$ is absorbing in $V_Z-y$
and $x\in Z$, so that $x-y=d\in Z-y$. Also $-d\in V_Z-y$ and because $Z-y$ is absorbing, there is $\alpha>0$ such that $-\alpha^{-1}d\in Z-y$, i.e.,\
$y-\alpha^{-1}d=z\in Z$ and
$$
y=\frac1{1+\alpha}x+\frac{\alpha}{1+\alpha}z.
$$
This means that for all $x\in Z$, $x\leq_Cy$, i.e.,\ $y\notin\partial Z$.
\qed

The weight function can be associated with a function called as Minkowski gauge. This connection gives more insight in the properties of the weight function in the
infinite-dimensional case. When $A$ is an absorbing subset of a real vector space $W$, we may define a function $P_A:W\to\mb R$,
$$
P_A(w)=\inf\{\alpha\geq0\,|\,\alpha^{-1}w\in A\},\quad w\in W.
$$
$P_A$ is called the {\it Minkowski gauge of $A$}. For basic properties of this function, we refer to \cite{infdimanalysis}. If $A$ is convex, then $P_A$ is a convex function, and
$$
\{v\in W\,|\,P_A(v)<1\}\subset A\subset\{v\in W\,|\,P_A(v)\leq1\}.
$$
When $A$ is an absorbing convex balanced subset, $P_A$ has many properties reminiscent to a norm, whose unit ball is $A$. When $W$ is a (locally convex) topological vector space,
the Minkowski gauge $P_A$ is continuous if and only if $A$ is a neighbourhood of the origin.

\begin{figure}[t]
\includegraphics[width=0.6\linewidth]{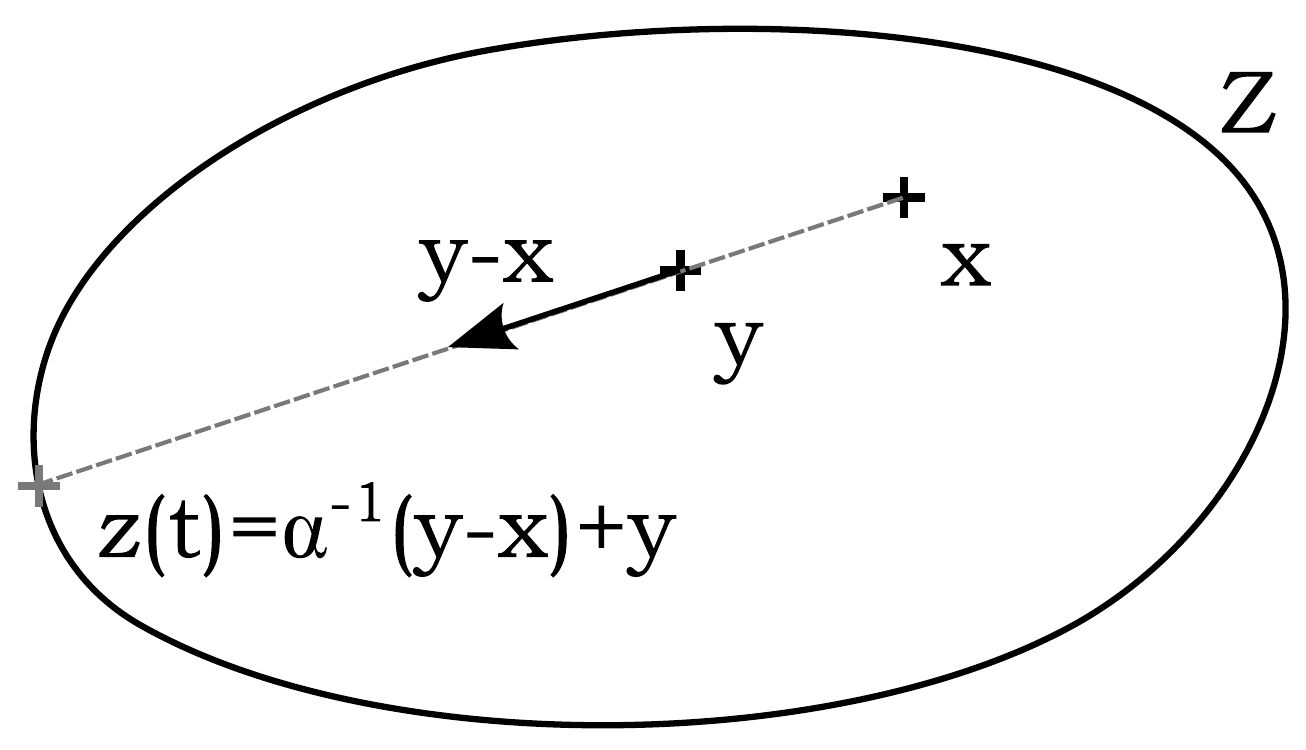}
\caption{%Illustration of the relation between Minkowski gauge $P_{\mc D(y)}(y-x)$ and the weight function $t_y(x)$.
The scalar $\alpha$ extending $y-x$ from the starting point $y$ to the boundary coincides with the Minkowski gauge $P_{Z-y}(y-x)$ and the decomposition of $y$ with respect
to this boundary point and $x$ gives the value $t_y(x)$ of the weight function.}
\label{fig:bdreltoMG}
\end{figure}

Suppose that $Z$ is a convex subset of a real vector space $V$ and $y\in Z$. The basis for connecting a Minkowski gauge to the weight function $t_y$ is provided by the following
observation: Consider a vector $y-x\in V_Z-y$, where $x\in Z$. As can be seen from Fig. (\ref{fig:bdreltoMG}), the scaling factor $\alpha$ that shrinks or extends this vector to the
border of the set $Z-y$ defines a point $z(t)$, which determines the value of the weight function $t_y$. These considerations can be formulated  mathematically as follows:
Pick $t\in[0,t_y(x))$ and define $z(t)=(1-t)^{-1}(y-tx)\in Z$. Now $z(t)-y=t(1-t)^{-1}(y-x)\in Z-y$. As $t$ approaches $t_y(x)$ from below, $\alpha(t)=(1-t)/t$ decreases and
from this we see that $\big(1-t_y(x)\big)/t_y(x)=P_{Z-y}(y-x)$ or, when we denote the Minkowski gauge $P_{Z-y}:V_Z-y\to[0,\infty)$ of $Z-y$ by $p_y(x)\equiv P_{Z-y}(y-x)$,
\begin{equation}\label{eq:t_yp_y}
t_y(x)=\frac1{1+p_y(x)}.
\end{equation}

According to Lemma \ref{prop:innerabsorb} the gauge $p_y$ is well defined, when $y\in Z\setminus\partial Z$. From the convexity of the Minkowski gauge we again see that
$x\mapsto1/t_y(x)=1+p_y(x)$ is convex on $Z$ whenever $y\in Z\setminus\partial Z$. We immediately see that, in the case of a topological vector space $V$, whenever
$y\in Z\setminus\partial Z$, the weight function $t_y$ is continuous if and only if the Minkowski gauge $p_y$ is continuous, i.e.,\ $Z-y$ is a neighbourhood of the origin of $V_Z-y$.
In finite-dimensional settings, any convex absorbing set is a neighbourhood of origin (as one may easily check). Thus we obtain the following result needed for proving Proposition
\ref{prop:extremal_deco}.

\begin{proposition}
Suppose that $Z\subset\mb R^n$ for some $n\in\mb N$. The weight function $t_y$ is continuous if and only if $y\in Z\setminus\partial Z$.
\end{proposition}

The quantum physical sets of states, POVMs and channels are all compact (even in the infinite-dimensional case with respect to suitable topologies), implying that, e.g.,\ Proposition
\ref{prop:extremal_deco} is applicable for the sets of (finite dimensional) quantum devices.

\section{Relation to Hilbert's projective metric}
\label{sec:apphpm}
The weight function is also related to the Hilbert's projective metric.
Suppose $C\subset V$ is a pointed generating
cone of a real vector space $V$ (see definition
in Section II).
We may define the functions
\begin{eqnarray*}
\inf(v/w)&=&\sup\{\lambda\in\mb R\,|\,v-\lambda w\in C\},\\
\sup(v/w)&=&\inf\{\lambda\in\mb R\,|\,\lambda w-v\in C\},
\end{eqnarray*}
$v,\,w\in V$. Through these functions, one can define {\it Hilbert's projective metric} $\mf h:V\times V\to[0,\infty]$,
%% $\mf h(v,w)=\ln{\big(\sup(v/w)\sup(w/v)\big)}$
$\mf h(v,w)=\ln{\big(\sup(v/w)/\inf(v/w)\big)}$
that can be
lifted into a well-defined metric in the projective space $\mb PV$; for more on this subject, see \cite{eveson1990,reeb_etal2011,gaubert2013}

When $Z$ is a base for $C$, one can easily show that, for $x,\,y\in Z$, $\inf(y/x)=\sup\{t\in[0,1)\,|\,y-tx\in C\}$. Moreover, if $x,\,y\in Z$ and $y-tx\in C$ for some $t\in[0,1)$, then $y-tx=sz$ for some (unique) $s\geq0$ and $z\in Z$. If $s\neq1-t$, then one sees that both $y\in Z$ and % $(s+t)^{-1}y\in Z$
\begin{align}
\frac{1}{s+t}\; y=\frac{t}{s+t}\; x+\frac{s}{s+t}\;z
\end{align}
belongs to $Z$
contradicting the fact that $Z$ is a base. Hence $s=1-t$ and
$$
\inf(y/x)=\sup\{t\in[0,1)\,|\,y-tx\in(1-t)Z\}=t_y(x).
$$
Similarly, the convex function $x\mapsto1/t_y(x)$ is associated with the $\sup$-function.

\section{Boundary of quantum convex sets}\label{sec:boundaryQCS}
The question of the boundary elements for states, observables and channels can be treated in a unified way as all these objects can be understood as transformations represented
by completely positive linear maps. In this section, we give conditions of being on the boundary for all relevant quantum devices. For the sake of brevity, we characterize the
boundary for all relevant quantum convex sets in one go. This, however, necessitates the use of Heisenberg picture which is
used only in this section.

Let us fix a Hilbert space $\mc H$ and a unital $C^* $-algebra $\mc A$. We say that a linear map $\Phi:\mc A\to\mc L(\hil)$ is {\it completely positive} (CP) if for any $n=1,\,2,\ldots$ and
$a_1,\ldots,\,a_n\in\mc A$ and $|v_1\ra,\ldots,\,|v_n\ra\in\hil$
$$
\sum_{j,k=1}^n\sis{v_j}{\Phi(a_j^\dagger  a_k)|v_k}\geq0.
$$
For any CP map $\Phi$ there is a Hilbert space $\mc M$, a linear map $J:\hil\to\mc M$ and a linear map $\pi:\mc A\to\mc L(\mc M)$ such that
$\pi(1)=I_{\mc M}$, $\pi(a^\dagger )=\pi(a)^\dagger $ and $\pi(ab)=\pi(a)\pi(b)$ for all $a,\,b\in\mc A$ (i.e.,\ $\pi$ is a unital *-representation of $\mc A$ on $\mc M$) that constitute a
{\it minimal Stinespring dilation} for $\Phi$. This means that $\Phi(a)=J^\dagger \pi(a)J$ for all $a\in\mc A$ and the subspace of $\mc M$ generated by the vectors $\pi(a)J|v\ra$, $a\in\mc A$,
$|v\ra\in\hil$, is dense in $\mc M$.

In what follows, we only study unital CP maps, i.e.,\ $\Phi(1_{\mc A})=I_\hil$. We denote the set of all unital CP maps $\Phi:\mc A\to\mc L(\hil)$ by $\mc{CP}(\mc A;\hil)$. Since the set
$\mc{CP}(\mc A;\hil)$ is convex, it is equipped with the preorder $\leq_C$. We denote $\Phi=_C\Psi$ if $\Phi\leq_C\Psi$ and $\Psi\leq_C\Phi$. For any $\Psi\in\mc{CP}(\mc A;\hil)$ we may define
the set
$$
\mc F(\Psi)=\{\Phi\in\mc{CP}(\mc A;\hil)\,|\,\Phi\leq_C\Psi\}.
$$
Let us fix a minimal dilation $(\mc M,\pi,J)$ for $\Psi$. Let us define $F(\Psi)$ as the set of positive operators $E\in\mc L(\mc M)$ such that $E\pi(a)=\pi(a)E$ for all $a\in\mc A$ and
$J^\dagger EJ=I$. The following proposition is essentially due to \cite{raginsky2003}.

\begin{proposition}\label{prop:FPsiEPsi}
Suppose that $\Psi\in\mc{CP}(\mc A;\hil)$ is equipped with the minimal dilation $(\mc M,\pi,J)$. The sets $\mc F(\Psi)$ and $F(\Psi)$ are in one-to-one correspondence set up by
\begin{equation}\label{eq:FPsiEPsi}
\Phi(a)=J^\dagger \pi(a)EJ,\qquad\Phi\in\mc F(\Psi),\quad E\in F(\Psi)
\end{equation}
for all $a\in\mc A$.
\end{proposition}

\begin{lemma}\label{lemma:PsieqPhi}
Suppose that $\Phi,\,\Psi\in\mc{CP}(\mc A;\hil)$ and fix the minimal dilation $(\mc M,\pi,J)$ for $\Psi$. Now $\Phi=_C\Psi$ if and only if there is $E\in F(\Psi)$ with bounded inverse such that
$\Phi(a)=J^\dagger \pi(a)EJ$ for all $a\in\mc A$.
\end{lemma}

\Proof
Case $\Phi=\Psi$ is obvious. Let us concentrate on the case $\Phi\neq\Psi$.

Let us assume that $\Phi=_C\Psi$. Because, especially, $\Phi\leq_C\Psi$, there is an operator $E\in F(\Psi)$ such that $\Phi(a)=J^\dagger \pi(a)EJ$ for all $a\in\mc A$. Denote the closure of the
range of $\sqrt E$ by $\mc M_E$ and the projection of $\mc M$ onto this subspace by $P_E$. Since $E$ commutes with $\pi$, also $P_E$ commutes with $\pi$, and we may define the map
$\pi_E:\mc A\to\mc L(\mc M_E)$, $\pi_E(a)=P_E\pi(a)|_{\mc M_E}$. Also define $J_E=\sqrt EJ$. It is straight-forward to check that the triple $(\mc M_E,\pi_E,J_E)$ constitutes a minimal dilation
of $\Phi$. Since also $\Psi\leq_C\Phi$ and $\Phi\neq\Psi$, it follows that there is $t\in(0,1)$ and $\Psi'\in\mc{CP}(\mc A;\hil)$ such that $\Phi=t\Psi+(1-t)\Psi'$. In other words, there is a number
$t\in(0,1)$ such that the map $\Psi'$,
$$
\Psi'(a)=\frac1{1-t}(\Phi-t\Psi)=\frac1{1-t}J^\dagger \pi(a)(E-tI)J,\qquad a\in\mc A,
$$
is completely positive or, equivalently, $E\geq tI$. Hence $E$ has a bounded inverse.

Suppose that $E\in F(\Psi)$ is as in the first part of the proof and $E^{-1}\in\mc L(\mc M)$. From Proposition \ref{prop:FPsiEPsi} it follows immediately that $\Phi\leq_C\Psi$. Denote
$E'=P_EE^{-1}|_{\mc M_E}$. We have $E'\geq0$, $J_E^\dagger E'J_E=J^\dagger J=I$ and
\begin{eqnarray*}
E'\pi_E(a)&=&P_EE^{-1}\pi(a)|_{\mc M_E}=P_EE^{-1}\pi(a)EE^{-1}|_{\mc M_E}\\
&=&P_EE^{-1}E\pi(a)E^{-1}|_{\mc M_E}=P_E\pi(a)E^{-1}|_{\mc M_E}\\
&=&\pi_E(a)E'
\end{eqnarray*}
for all $a\in\mc A$, so that $E'\in F(\Phi)$ when we fix the dilation $(\mc M_E,\pi_E,J_E)$ for $\Phi$. Furthermore
$$
J_E^\dagger \pi_E(a)E'J_E=J^\dagger \pi(a)\sqrt EE^{-1}\sqrt EJ=J^\dagger \pi(a)J=\Psi(a)
$$
for all $a\in\mc A$. According to Proposition \ref{prop:FPsiEPsi} this means that $\Psi\leq_C\Phi$.
\qed

We denote the spectrum of an operator $E\in\mc L(\mc M)$ on a Hilbert space $\mc M$ by $\mr{sp}(E)$. The following proposition, which is an immediate corollary of the previous lemma,
characterizes the boundary elements of the set of unital CP maps.

\begin{proposition}\label{prop:boundaryCP}
Suppose that $\Phi\in\mc{CP}(\mc A;\hil)$. The map $\Phi$ is on the boundary of $\mc{CP}(\mc A;\hil)$ if and only if there is $\Psi\in\mc{CP}(\mc A;\hil)$ with a minimal dilation
$(\mc M,\pi,J)$ such that $\Phi\leq_C\Psi$ and $\Phi$ corresponds to an operator $E\in F(\Psi)$ with $0\in\mr{sp}(E)$.
\end{proposition}

\Proof
The condition $\Phi\in\partial\mc{CP}(\mc A;\hil)$ is equivalent with the fact that there is $\Psi\in\mc{CP}(\mc A;\hil)$ such that $\Phi\leq_C\Psi$ but $\Psi\not\leq_C\Phi$. Indeed, if
$\Psi'\in\mc{CP}(\mc A;\hil)$ is such that $\Psi'\not\leq_C\Phi$, we may define $\Psi=\frac12\Phi+\frac12\Psi'$ so that $\Phi\leq_C\Psi$. Moreover, if $\Psi\leq_C\Phi$, it would follow that
$\Psi'\leq_C\Psi\leq_C\Phi$ yielding $\Psi'\leq\Phi$ yielding a contradiction. Suppose that $\Psi\in\mc{CP}(\mc A;\hil)$ is such that $\Phi\leq_C\Psi$ and $\Psi\not\leq_C\Phi$ and $\Psi$ has the
minimal dilation $(\mc M,\pi,J)$ and $\Phi$ corresponds to the operator $E\in F(\Psi)$ according to Equation (\ref{eq:FPsiEPsi}). According to Lemma \ref{lemma:PsieqPhi}, the condition
$\Psi\not\leq_C\Phi$ is equivalent to $E$ not having a bounded inverse or, in other words, $0\in\mr{sp}(E)$.
\qed

The CP maps of quantum physics are normal. This is because in this section we have described our quantum devices jointly in Heisenberg picture and, in order to transcend to the Schr\"odinger
picture, we generally need normality. However, when $\hil$ and $\mc A$ are finite dimensional the maps $\Phi\in\mc{CP}(\mc A;\hil)$ are automatically normal. The results of this section also
hold for the restricted class of normal elements in $\mc{CP}(\mc A;\hil)$ because this class is a face of $\mc{CP}(\mc A;\hil)$, i.e.,\ if $\Phi$ is normal and $\Phi'\leq_C\Phi$ then also $\Phi'$ is
normal.

\subsection{States}
\label{sec:appbfstates}

Suppose that $\mc K$ is a Hilbert space. We will denote the set of states by $\mc S(\mc K)$ containing positive trace-class operators on $\mc K$ with trace 1. The states are in one-to-one
correspondence with the normal (completely) positive unital maps $\f:\mc L(\mc K)\to\mb C$, i.e. the set of normal elements in $\mc{CP}\big(\mc L(\mc K);\mb C\big)$.

\begin{proposition}\label{prop:boundarystate}
A state $\varrho\in\partial\mc S(\mc K)$ if and only if $\rho$ has 0 in its spectrum.
\end{proposition}

\Proof
First, let us assume that $\dim{\mc K}<\infty$. Suppose that $\varrho\in\mc S(\mc K)$ is such that there is a unit vector $|v\ra\in\mc K$ such that $\varrho|v\ra=0$. Let us define the operator
$D=\ktb{v}{v}-(d-1)^{-1}(I-\ktb{v}{v})$. Denote the smallest non-zero eigenvalue of $\varrho$ by $\lambda_{min}$. It is easy to see that whenever $\eps\leq\lambda_{\min}$,
$\varrho+\eps D\in\mc S(\mc K)$ but $\varrho-\eps D$ is not positive for any $\eps>0$. Hence $\varrho\in\partial\mc S(\mc K)$.

Suppose now $\varrho\in\partial\mc S(\mc K)$, i.e.,\ there is a state $\sigma\in\mc S(\mc K)$ such that when we denote $D=\sigma-\varrho$, then $\varrho-\eps D$ is not positive for any
$\eps>0$. We may write $\mc K=\mc K_+\oplus\mc K_0\oplus\mc K_-$, where $\mc K_+$ is the direct sum of the eigenspaces corresponding to the positive eigenvalues of $D$ and $\mc K_0$ is
the kernel of $D$. We infer that $\mc K_+\cap\ker{\varrho}$ is non-trivial and hence also $\ker{\varrho}$ is non-trivial. This means that $0$ is an eigenvalue of $\varrho$.

Now, let us assume that $\mc K$ is infinite dimensional. Assume that $\varrho\in\mc S(\mc K)$ would be in the interior, i.e.,\ $\varrho\notin\partial\mc S(\mc K)$. Then, especially,
$\ktb{v}{v}\leq_C\varrho$ for all unit vectors $|v\ra\in\mc K$. Whenever $\lambda\ktb{v}{v}\leq A$ for some $\lambda>0$ and some positive $A\in\mc L(\mc K)$, it follows \cite{buschgudder}
that $|v\ra\in\mr{ran}(\sqrt A)$ or, in other words, $|v\ra=\sqrt A|w\ra$ for some $|w\ra\in\mc K$. In the case where $A$ is a state operator, this result was already proven in
\cite{hadjisavvas}. Hence, $\mr{ran}(\sqrt\varrho)=\mc K$, i.e.,\ $\sqrt\varrho$ is surjective. If $\varrho$ had a non-trivial kernel, it could not be in the interior for then
$\ktb{v}{v}\not\leq_C\varrho$ for any unit vector $|v\ra\in\mr{ker}(\varrho)$. Hence, $\rho$ is injective and so $\sqrt\varrho$ is injective as well. All this implies that
$\sqrt\varrho:\mc K\to\mc K$ is a bijection and the open mapping theorem yields that there is a continuous inverse $\sqrt\varrho^{-1}:\mc K\to\mc K$. Hence, there is a bounded inverse
$\varrho^{-1}=\sqrt\varrho^{-1}\sqrt\varrho^{-1}$. However, this is impossible, since in the infinite-dimensional case all state operators have 0 in their spectra.
\qed

The previous proposition tells us that the boundary of the set of states depends dramatically on the dimensionality of the Hilbert space: If the space is finite dimensional, boundary states are
exactly those whose kernel is non-trivial. In the infinite-dimensional case, the set of states coincides with its boundary.

\subsection{Effects and finite outcome observables}\label{sec:appbfobs}

Denote $\Om=\{1,\ldots,\,N\}$ and define $\mc O^N(\hil)$ as the set of {\it positive-operator-valued measures on $\hil$ and taking values in $\Om$} ($N$-outcome observables), i.e.,
$\ms M\in\mc O^N(\hil)$ is a collection $\ms M=\{M_j\}_{j=1}^N$ of positive operators on $\hil$ such that $\sum_{j=1}^NM_j=I$. It should be noted that whenever $\ms M\in\mc O^N(\hil)$
then $\ms M\leq_C\ms E^N$, where $\ms E^N=\{E_j^N\}_{j=1}^N$, $E_j^N=N^{-1}I$ for all $j=1,\ldots,\,N$. Note that we may identify $\mc O^N(\hil)$ with the set of normal elements in
$\mc{CP}(\mc A^N,\hil)$, where $\mc A^N$ is just the algebra $\mb C^N$ with componentwise operations.

\begin{proposition}\label{prop:boundaryeffect}
The boundary $\partial\mc O^N(\hil)$ consists of POVMs $\ms M=\{M_j\}$ with $0\in\mr{sp}(\ms M_j)$ for some $j=1,\ldots,\,N$.
\end{proposition}

\Proof
Endow $\mb C^N$ with an orthonormal basis $\{|1\ra,\ldots\,|N\ra\}$ and denote $P_r=\ktb{r}{r}$, $r=1,\ldots\,N$. Define the PVM $\ms Q\in\mc O^N(\hil\otimes\mb C^N)$,
$Q_r=I\otimes P_r$, $r=1,\ldots\,N$, and the isometry $J:\hil\to\hil\otimes\mb C^N$, $J|v\ra=N^{-1/2}|v\ra\otimes(|1\ra+\cdots+|N\ra)$. It is immediately seen that
$(\hil\otimes\mb C^N,\ms Q,J)$ is a minimal dilation of $\ms E^N=\{N^{-1}I,\ldots\,N^{-1}I\}$, i.e.,\ $J^\dagger Q_rJ=N^{-1}I$, $r=1,\ldots\,2$. Let $F(\ms E^N)$ be the set of positive
operators $E$ on $\hil\otimes\mb C^N$ that commute with $\ms Q$ and $J^\dagger EJ=I$ so that $\mc O^N(\hil)$ is in one-to-one affine correspondence with $F(\ms E^N)$. It follows that
$F(\ms E^N)$ consists of operators of the form $\sum_{j=1}^NA_j\otimes P_j$, where $A_j\in\mc L(\hil)$ are positive operators with $A_j\leq2I$. Any $\ms M\in\mc O^N(\hil)$ corresponds to
such an operator, where $A_j=2M_j$. A POVM $\ms M$ is thus on the boundary if and only if the corresponding operator $2\sum_{j=1}^NM_j\otimes P_j$ has 0 in its spectrum. This happens
exactly when $0\in\mr{sp}(M_j)$ for some $j$.
\qed

It is often denoted $\mc O^2(\hil)=\mc E(\hil)$ and $\ms E\in\mc E(\hil)$ are called {\it effects}. An effect $\ms E=\{E_1,E_2\}\in\mc E(\hil)$ is usually identified with its value $E_1$ and hence
effects are characterized as positive operators $E\in\mc L(\hil)$ with $E\leq I$. One easily sees from the previous proposition that an effect $E$ is on the boundary if and only if $0\in\mr{sp}(E)$
or $1\in\mr{sp}(E)$.

\subsection{Channels}
\label{sec:appbfchannels}

In this subsection, we assume that $\hil$ and $\mc K$ are (separable) Hilbert spaces. We denote by $\mc C(\mc K;\hil)$ the set of (normal) unital CP maps $\mc E:\mc L(\mc K)\to\mc L(\hil)$
and call these maps as {\it channels}. Note the the physical input space of these channels is $\hil$ and output is $\mc K$. The minimal Stinespring dilation $(\mc M,\pi,J)$ of a channel
$\mc E\in\mc C(\mc K;\hil)$ can be chosen so that $\mc M$ is separable and $\pi:\mc L(\mc K)\to\mc L(\hil)$ is a normal unital *-representation. This means that there is a separable Hilbert
space $\mc K'$ such that we may choose $\mc M=\mc K\otimes\mc K'$ and $\pi(B)=B\otimes I_{\mc K'}$ for all $B\in\mc L(\mc K)$. Hence we usually denote a minimal Stinespring dilation of a
channel $\mc E$ in the form $(\mc K',J)$ where $J:\hil\to\mc K\otimes\mc K'$ is an isometry such that
$$
\mc E(B)=J^\dagger (B\otimes I_{\mc K'})J,\qquad B\in\mc L(\mc K).
$$

Suppose that $\mc K$ is infinite-dimensional and $\mc E\in\mc C(\mc K;\hil)\setminus\partial\mc C(\mc K;\hil)$. For each unit vector $|v\ra\in\mc K$ define the channel
$\mc F^{|v\ra}\in\mc C(\mc K;\hil)$ by $\mc F^{|v\ra}(B)=\sis{\f}{B\f}I$. The predual map $\mc F_*^{|v\ra}:\mc T(\hil)\to\mc T(\mc K)$ of $\mc F^{|v\ra}$ is given by
$\mc F_*^{|v\ra}(T)=\tr{T}\ktb{v}{v}$ for all trace-class operators $T\in\mc T(\hil)$. It follows that $\mc F^{|v\ra}\leq_C\mc E$ for all unit vectors $|v\ra\in\mc K$ which means that for all
unit vectors $|v\ra\in\mc K$ there is a number $t_{|v\ra}\in(0,1]$ such that for all positive $T\in\mc T(\hil)$ and $B\in\mc L(\mc K)$ one has
$$
\tr{T(\mc E-t_{|v\ra}\mc F^{|v\ra})(B)}=\tr{B(\mc E_*-t_{|v\ra}\mc F_*^{|v\ra})(T)}\geq0
$$
yielding $t_{|v\ra}\mc F_*^{|v\ra}(T)\leq\mc E_*(T)$. By picking a positive operator $T$ of trace one, we find that $\ktb{v}{v}\leq_C\mc E_*(T)$ for all unit vectors $|v\ra\in\mc K$ when
$\mc E_*(T)$ is considered as a state. As in the proof of Proposition \ref{prop:boundarystate}, one can show that this result leads into a contradiction. This means that if $\mc K$ is infinite
dimensional, $\mc C(\mc K;\hil)$ coincides with its boundary.

Suppose that $\dim{\mc K}=d<\infty$ and fix an orthonormal basis $\{|n\ra\}_{n=1}^d\subset\mc K$. Define for each $\mc F\in\mc C(\mc K;\hil)$ the {\it Choi operator}
$$
E(\mc F)=d\sum_{m,n=1}^d\ktb{m}{n}\otimes\mc F(\ktb{m}{n})\in\mc L(\mc K\otimes\hil).
$$
Define the vector $|\psi_d\ra=d^{-1/2}(|1,1\ra+\cdots+|d,d\ra)\in\mc K\otimes\mc K$ and the isometry $J:\hil\to\mc K\otimes\mc K\otimes\hil$ with $J|v\ra=|\psi_d\ra\otimes|v\ra$ for all
$|v\ra\in\hil$. One can easily check that the pair $(\mc K\otimes\hil,J)$ constitutes a minimal dilation for the channel $\mc E\in\mc C(\mc K;\hil)$, $\mc E(B)=d^{-1}\tr{B}I_\hil$. Suppose that
$\mc F\in\mc C(\mc K;\hil)$. We find
\begin{eqnarray*}
J^\dagger \big(B\otimes E(\mc F)\big)J&=&d\sum_{m,n=1}^d\sis{\psi_d}{B\otimes\ktb{m}{n}|\psi_d}\mc F(\ktb{m}{n})\\
&=&\sum_{m,n,r,s=1}^d\sis{r}{B|s}\sis{r}{m}\sis{n}{s}\mc F(\ktb{m}{n})\\
&=&\sum_{m,n=1}^d\sis{m}{B|n}\mc F(\ktb{m}{n})=\mc F(B)
\end{eqnarray*}
for all $B\in\mc L(\mc K)$. This means that $\mc C(\mc K;\hil)=\mc F(\mc E)$ when $\mc K$ is finite-dimensional and the operator on the dilation space of $\mc E$ corresponding to a channel
$\mc F\in\mc C(\mc K;\hil)$ is the Choi operator. Hence we can give the following characterization for boundary channels:

\begin{proposition}\label{prop:boundarychannelfinite}
Suppose that $\dim{(\mc K)}<\infty$. A channel $\mc F\in\mc C(\mc K;\hil)$ is on the boundary $\partial\mc C(\mc K;\hil)$ if and only if the Choi operator $E(\mc F)$ has 0 in its spectrum.
\end{proposition}

In the case when both $\dim{\mc K}=d_{\mc K}$ and $\dim\hil=d_\hil$ are finite, the above result means that a channel is on the boundary if and only if its Kraus rank is strictly less than
$d_{\mc K}d_\hil$. Suppose now that $\{|m\ra\}_{m=1}^{d_\hil}\subset\hil$ is an orthonormal basis. Since $E(\mc F)$ is positive for any channel $\mc F$, we may give it the
spectral decomposition $E(\mc F)=d_{\mc K}\sum_{j=1}^r\ktb{L_j}{L_j}$. Let us define the operators
$L_j=\sum_{m=1}^{d_\hil}\sum_{n=1}^{d_{\mc K}}\sis{n,m}{L_j}\ktb{m}{n}$. One may check that the operators $K_j=L_j^\dag$ constitute a minimal set of Kraus operators for
$\mc F$, i.e.,\ $\mc F(B)=\sum_{j=1}^rK_j^\dag BK_j$. Moreover, the more familiar Choi operator associated with the Schr\"odinger (predual) version of $\mc F$ is given by
$$
C(\mc F_*)=\sum_{m,n=1}^{d_\hil}\ktb{m}{n}\otimes\mc F_*(\ktb{m}{n})=\sum_{j=1}^r\ktb{K_j}{K_j},
$$
where  %%is the operator $K_j$ expressed through the Choi-Jamiolkowski isomorphism.
$|K_j\ra=\sum_{m=1}^{d_\hil}\sum_{n=1}^{d_{\mc K}} \bra{n}K_j\ket{m} \ket{m,n}=\sum_{m=1}^{d_\hil}\sum_{n=1}^{d_{\mc K}} \bra{L_j}n,m\ra\ \ket{m,n}$,
$|K_j\ra\in\hil\otimes\mc K$.
Let us note, that orthogonality of vectors $\ket{L_j}$ implies the orthogonality of vectors $\ket{K_j}$, while their norm $\bra{L_j}L_j\ra^{1/2}=\bra{K_j}K_j\ra^{1/2}$ is the same.
Hence, we demonstrated the following.

\begin{proposition}\label{prop:schrodinger}
Suppose that $\dim{\mc K}=d_{\mc K}$ and $\dim\hil=d_\hil$ are finite. A completely positive trace preserving map (i.e. a channel in the Schr\"odinger picture)
%$\mc F_*\in\mc C(\hil;\mc K)$
is on the boundary of the set of channels
%$\partial\mc C(\mc K;\hil)$
if and only if the rank of its Choi operator is strictly less than $d_{\mc K}d_\hil$.
\end{proposition}

Thus, also in the Schr\"odinger picture the channel is on the boundary, when zero is the spectrum of its Choi operator.

\section{Evaluation of boundariness for a qubit ``erasure'' channel}
\label{sec:appccchannel}
The aim of this appendix is to study two-element convex decompositions of the channel $\cE_p$ into extremal rank-two qubit channels $\mathcal{F}$ and channels $\mathcal{G}$.
Any such channel $\mathcal{F}$ has a Choi matrix, which can be written in the spectral form:
\begin{align}
\label{eq:ccdeff}
F=\frac{1}{2}(1+q)\ket{\psi}\bra{\psi}+\frac{1}{2}(1-q)\ket{\phi}\bra{\phi}\,,
\end{align}
where $\ket{\psi},\ket{\phi}$ are mutually orthogonal unit vectors
on $\mc{H}_2\otimes\mc{H}_2$ and $0\leq q < 1$, hence $\tr{F}=1$.
%%As in the main text,
Vectors $\ket{\psi}$,$\ket{\phi}$ can be written in the Schmidt form
\begin{align}
\label{eq:ccschmidt}
\ket{\psi}&=\sqrt{s}\ket{u}\ket{v}+\sqrt{1-s}\ket{u_\perp}\ket{v_\perp}, \\
\ket{\phi}&=\sqrt{r}\ket{w}\ket{v^\prime}+\sqrt{1-r} \ket{w_\perp}\ket{v^\prime_\perp}
\end{align}
with $1/2< s\leq 1$ and
$0 \leq r\leq 1$. Let us note that $s=1/2$ does not correspond to an extremal channel, but to a mixture of unitary channels (i.e. it leads to $r=1/2$).
The condition ${\rm tr}_1 F = \frac{1}{2}I$
requires that $\ket{v^\prime}=\ket{v}$ and %%$r=\frac{1-(1+q)s}{1-q}$.
\begin{align}
\label{eq:relqsr}
r=\frac{1-(1+q)s}{1-q}.
\end{align}
The orthogonality $\la{\psi}|{\phi}\ra=0$ gives
\begin{align}
\label{eq:ccortho}
0=\sqrt{sr}\la{u}|{w}\ra+\sqrt{(1-s)(1-r)}\la{u_\perp}|{w_\perp}\ra .
\end{align}
For any two states of a qubit it holds that $|\la{u}|{w}\ra|=|\la{u_\perp}|{w_\perp}\ra |$.
Thus, Eq. (\ref{eq:ccortho}) can be satisfied only in two ways: i) $\la{u}|{w}\ra=-\la{u_\perp}|{w_\perp}\ra$ and $rs=(1-s)(1-r)$, which is, according to Eq. (\ref{eq:relqsr}),
equivalent to $q=0$ ii) both overlaps in Eq. (\ref{eq:ccortho}) vanish.

Let us start with the case i),
i.e both nonzero eigenvalues of Choi operator $F$ are equal to $1/2$
and the scalar products of vectors $u,v$ and $u_\perp, v_\perp$ have
opposite sign.
Since channel $\mathcal{G}$ must belong to the boundary of the set of channels, there exists a normalized vector $\ket{\varphi}$ from the kernel of $G$, i.e. $\bra{\varphi}G\ket{\varphi}=0$.
We compute the expectation value of $E_p$ along the vector $\ket{\varphi}$. Using Eq. (\ref{eq:ccdecomp1}) we get
\begin{align}
\label{eq:chr2lb1}
\frac{p}{2}\leq \bra{\varphi}E_p\ket{\varphi}=t \bra{\varphi}F\ket{\varphi}=t c,
\end{align}
where the lower bound on the left follows from the eigenvalues of $E_p$ being greater or equal to $p/2$ and we denoted $c\equiv\bra{\varphi}F\ket{\varphi}$.
We notice that $0<c \leq 1/2$, because $F$ is positive semidefinite and its eigenvalues are zero and $1/2$.
From Eq. (\ref{eq:chr2lb1}) we get the lower bound $t\geq p/(2c) \geq p >p(1-p)$.
In other words the weight function $t_{\cE_p}(\mathcal{F})$ gives on these channels $\mathcal{F}$ values higher then $p(1-p)$.
Thus, we conclude that the convex decompositions (\ref{eq:ccdecomp1}) with rank-two
channels $\mathcal{F}$ having $\la{u}|{w}\ra\neq 0$ can not achieve as small value of $t$ as it is achieved by the unitary channels.

So let us investigate the case ii) and assume $\la{u}|{w}\ra=0$.
Our aim is to show that also in this case $t>p(1-p)$.
Unfortunately, we were
not able to solve this part of the problem completely analytically and we had to rely on numerical approach outlined in Remark \ref{prop:algorithm}.
Thus, the test whether the Choi-operators $G$ generated by operators $F$ and the weight $p(1-p)$ correspond to channels was done numerically.
More precisely, for $t=p(1-p)$ we calculated the smallest eigenvalues of $G$
for many choices of $F$ from the current subclass of extremal rank-two qubit
channels and we confirmed that in all cases the obtained value is non-negative, i.e. $G$ always corresponded to a channel. Below are some details on how the actual test was done.

Without loss of generality we can write
\begin{align}
\label{eq:ccmc1}
\ket{\phi}&=\sqrt{r}\ket{u_\perp}\ket{v}+e^{i\alpha}\sqrt{1-r} \ket{u}\ket{v_\perp} .
\end{align}
The Choi-operator $E_p$ is invariant under the unitary transformations $I\otimes V$ on the input Hilbert space. These transformations do not change eigenvalues, so to investigate
eigenvalues of $G$ we can equivalently investigate $(I\otimes V) \, G \, (I\otimes V^\dagger)$, which is for $V\ket{v}=\ket{0}$ the same as choosing $\ket{v}=\ket{0}$ in Eqs.
(\ref{eq:ccschmidt}), (\ref{eq:ccmc1}) and working directly with $G$. Moreover, we parameterize the vectors $\ket{u}$, $\ket{u_\perp}$ as:
\begin{align}
\label{eq:ccdefog}
&\ket{u}=\cos{\frac{\theta}{2}}\ket{0}+e^{i\beta}\sin{\frac{\theta}{2}} \ket{1} \nonumber \\
&\ket{u_\perp}=e^{i\gamma}\sin{\frac{\theta}{2}}\ket{0}-e^{i(\gamma+\beta)} \cos{\frac{\theta}{2}}\ket{1}
\end{align}
In this way operator
\begin{align}
G&=\frac{1}{1-p(1-p)} \left(E_p - p(1-p)F \right)
\end{align}
further specified by Eqs.
(\ref{eq:ccdeff}-\ref{eq:ccschmidt}), (\ref{eq:relqsr}), (\ref{eq:ccmc1}-\ref{eq:ccdefog})
and $\ket{v}=\ket{0}$ becomes a function of parameters $q,s,\alpha,\beta,\gamma,\theta$.
Let us note that Eq. (\ref{eq:relqsr}) requires parameters $q$ and $s$ to fulfill $s \leq 1/(1+q)$, since one must have $r\geq 0$. Especially, $q\rightarrow 1$ requires
$s\rightarrow 1/2$ and the operator $F$ converges to a Choi operator of a unitary channel. In such case we expect that $\lambda_G$, the minimal eigenvalue of $G$, will converge
to zero, because $\mathcal{G}$ must converge to a boundary in the set of channels.

\begin{figure}[t]
 \centering
\subfigure{%
\includegraphics[width=0.48\linewidth]{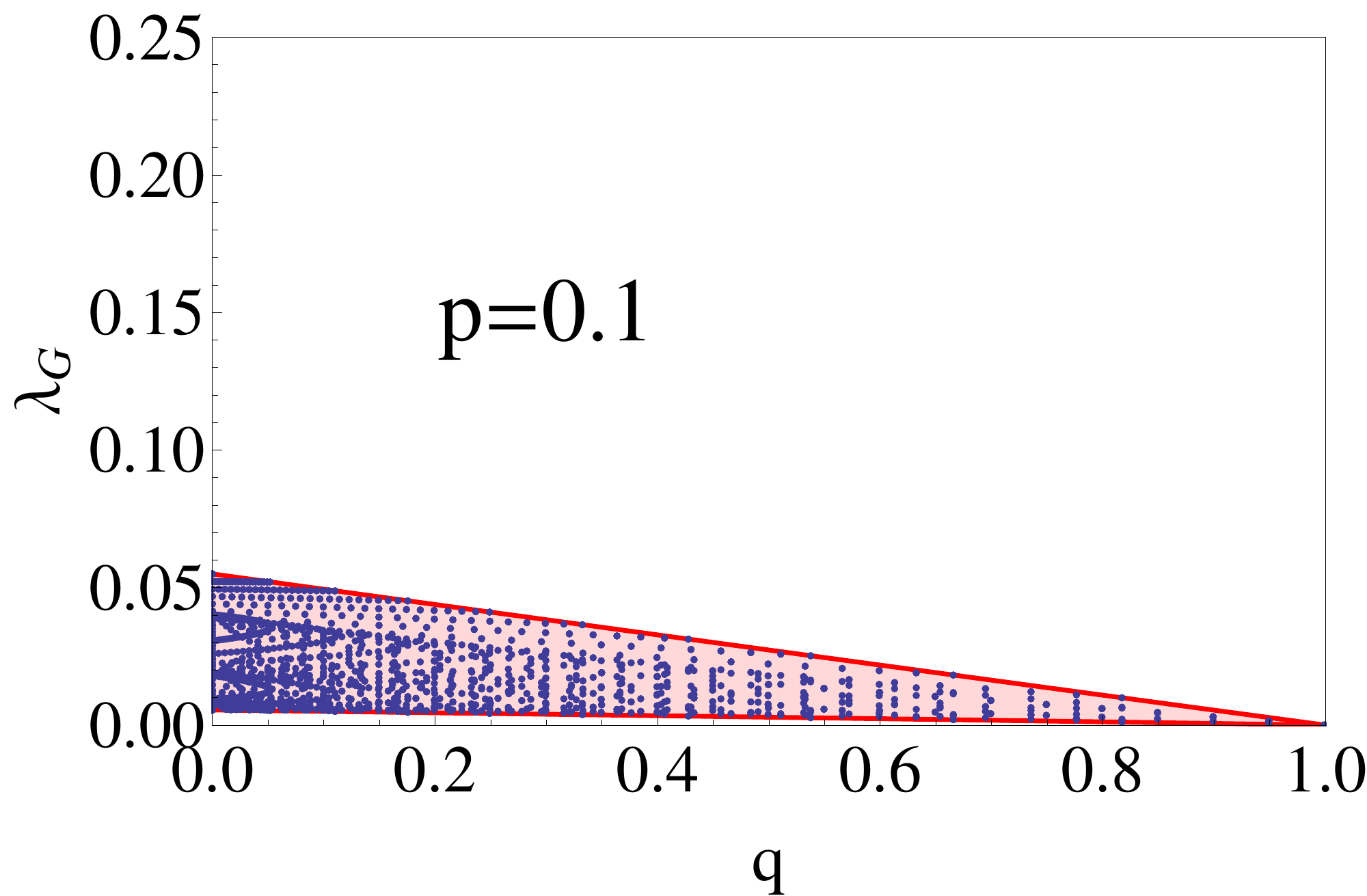}
\label{fig:subfigure1}}
\subfigure{%
\includegraphics[width=0.48\linewidth]{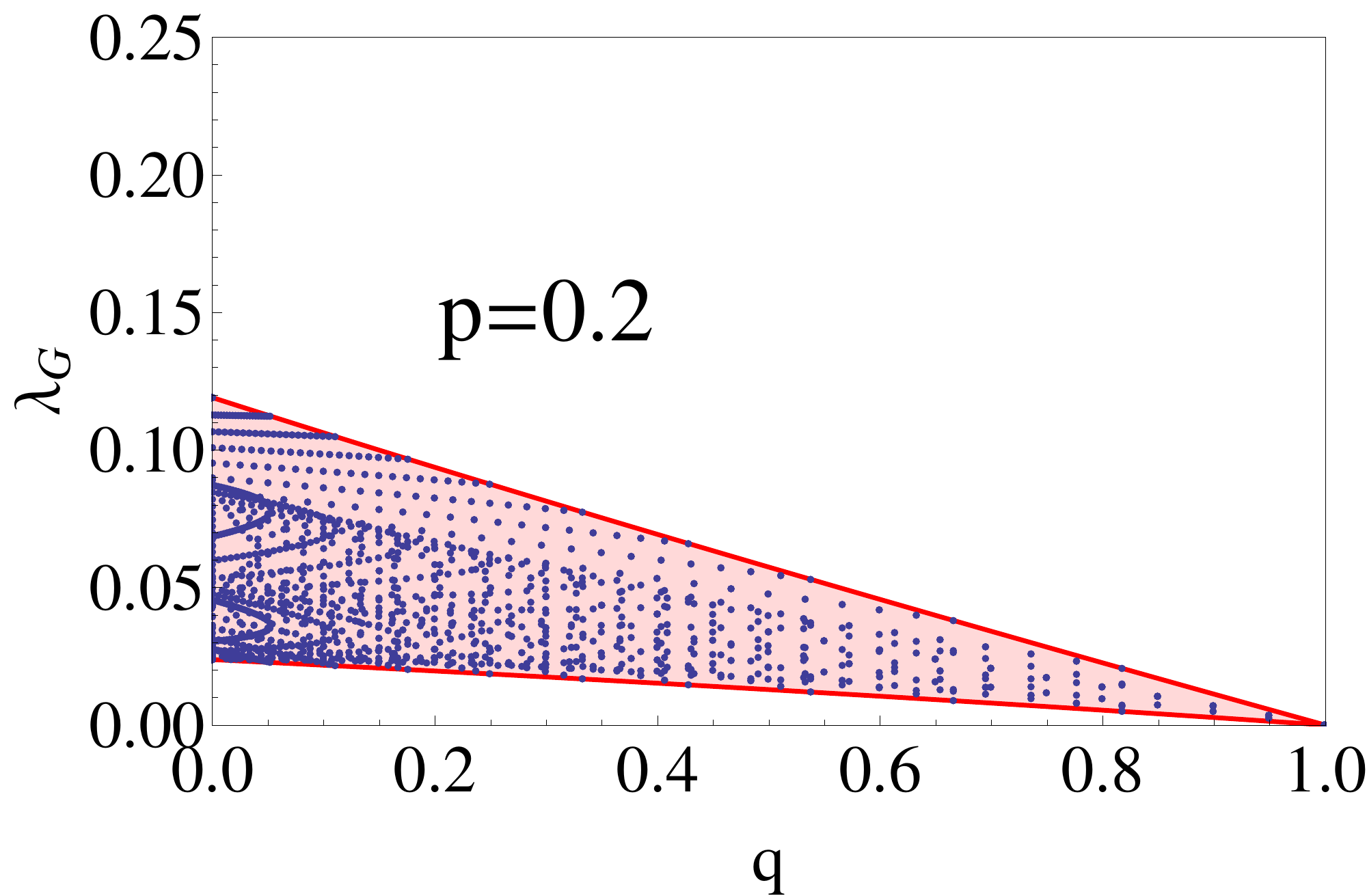}
\label{fig:subfigure2}}\\
\subfigure{%
\includegraphics[width=0.48\linewidth]{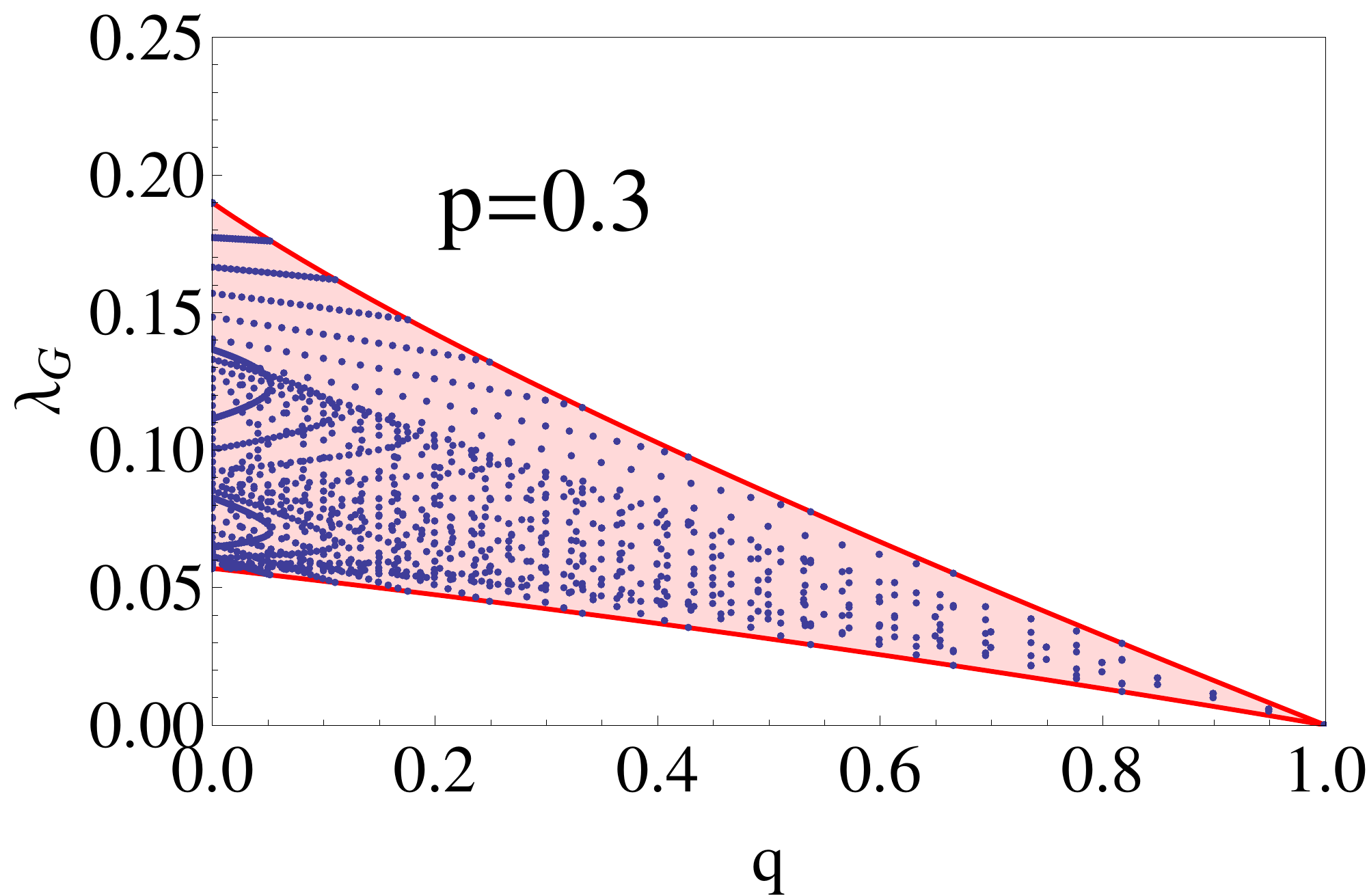}
\label{fig:subfigure3}}
\subfigure{%
\includegraphics[width=0.48\linewidth]{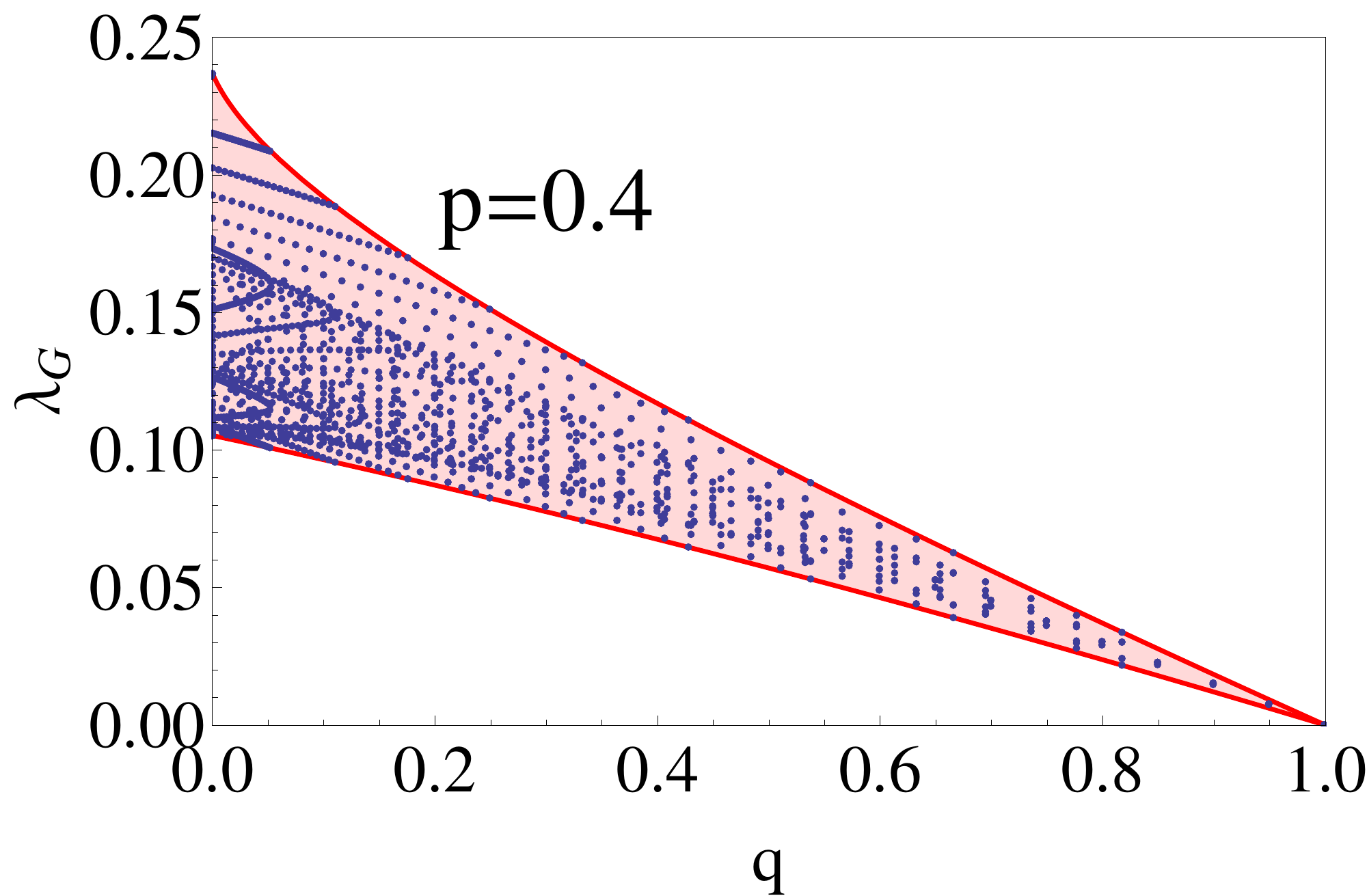}
\label{fig:subfigure4}}
\caption{(Color online) Illustration of the dependance of the minimal eigenvalue $\lambda_G$ of operator $G$ on the parameter $q$ depicted for different values of the remaining parameters $s,\alpha,\beta,\gamma,\theta$ and $p$.}
\label{fig:numresults}
\end{figure}

For this reason it is useful to plot $\lambda_G$ as a function of $q$ for some choice of remaining parameters (see Fig. \ref{fig:numresults}). By numerically analyzing the actual
dependence of the graphs on the parameters $s,\alpha,\beta,\gamma,\theta$ we observed that for a fixed $q$ the minimum and the maximum value of $\lambda_G$ can be achieved
only when $s=1/(1+q)$ and $\theta=0$; $\theta=\pi/2$, respectively. In such case parameters $\alpha,\beta$ and $\gamma$ do not influence $\lambda_G$
and it can be calculated analytically. The obtained dependencies $G_{min}(p,q)$ and $G_{max}(p,q)$ are visualized on Fig. \ref{fig:numresults} as red lines, which form the boundary
of the area where $\lambda_G$, the minimal eigenvalue of $G$, lies for any possible choice of its parameters. We can show that the minimum of $G_{min}(p,q)$ is zero and it is
achieved only for $q=1$ corresponding to a unitary channel $\mathcal{F}$. Similarly, all the blue points in the Fig. \ref{fig:numresults} corresponding to the minimal eigenvalue of $G$
for some choice of its parameters were having $\lambda_G>0$, which proves that $G\geq 0$ in the considered range of parameters $q,s,\alpha,\beta,\gamma,\theta$. In conclusion, we proved that the boundariness is indeed achieved for
decompositions containing at least one unitary channel, thus, it reads
$b(\cE_p)=p(1-p)$.

%%%%%%%%%%%%%%%%%%%%%%%%%%%%%%%%%%%%%%%%%%%%%%%%%%%%%%%%%%%%%%%%%%%%%%%%%%
%%%%%%%%%%%%%%%%%%%%%%%%%%%%%%%%%%%%%%%%%%%%%%%%%%%%%%%%%%%%%%%%%%%%%%%%%%
%%%%%                      BIBLIOGRAPHY                             %%%%%%
%%%%%%%%%%%%%%%%%%%%%%%%%%%%%%%%%%%%%%%%%%%%%%%%%%%%%%%%%%%%%%%%%%%%%%%%%%
%%%%%%%%%%%%%%%%%%%%%%%%%%%%%%%%%%%%%%%%%%%%%%%%%%%%%%%%%%%%%%%%%%%%%%%%%%

%%%%%%%%%%%%%%%%%%%%%%%%%%%%%%%%%%%%%%%%%%%%%%%%%%%%%%%%%%%%%%%%%%%%%%%
%%%%%%%%%%%%%%%%%%%%%%%%%%%%%%%%%%%%%%%%%%%%%%%%%%%%%%%%%%%%%%%%%%%%%%%
%%%%                     END OF DOCUMENT                           %%%%
%%%%%%%%%%%%%%%%%%%%%%%%%%%%%%%%%%%%%%%%%%%%%%%%%%%%%%%%%%%%%%%%%%%%%%%
%%%%%%%%%%%%%%%%%%%%%%%%%%%%%%%%%%%%%%%%%%%%%%%%%%%%%%%%%%%%%%%%%%%%%%%

\end{document}